\documentclass[]{aa}  

\usepackage{graphicx}
\usepackage{amssymb}
\usepackage{amsmath}
\usepackage{color}
\usepackage{subcaption}
\usepackage{appendix}
\usepackage{caption}
\usepackage{tabularx}
\usepackage{graphicx,url,twoopt,natbib}
\usepackage[varg]{txfonts}
\usepackage{enumerate}
\usepackage{ulem,soul}
\usepackage{threeparttable}
\usepackage{CJKutf8}
\bibpunct{(}{)}{;}{a}{}{,} 
\usepackage[colorlinks,
            linkcolor=red,
            anchorcolor=blue,
            citecolor=blue]{hyperref}

\begin{document} 
\begin{CJK}{UTF8}{gbsn}
% \linenumbers
   \title{Surface brightness-color relations for red giant branch stars using asteroseismic radii and Gaia distances (the ARD method)}
   
   \author{Jianping Xiong$^{\dag}$
          \inst{1,2,3},
          Qiyuan Cheng
          \inst{1,2,3,5},
          Xiaodian Chen
          \inst{4,5,6},
          Jiao Li
          \inst{1,2,3},
          Xiaobin Zhang
          \inst{4,5,6},
          Zhanwen Han
          \inst{1,2,3,5},
          \and
          Xuefei Chen$^{\dag}$\inst{1,2,3,5}
          }

   \institute{Yunnan Observatories, Chinese Academy of Sciences, 396 YangFangWang, Guandu District, Kunming, 650216, Peopleʼs Republic of China\\
              \email{cxf@ynao.ac.cn, xiongjianping@ynao.ac.cn}
         \and
             Key Laboratory for Structure and Evolution of Celestial Objects, Chinese Academy of Sciences, P.O. Box 110, Kunming 650216, People's Republic of China
        \and
             International Centre of Supernovae, Yunnan Key Laboratory, Kunming 650216, People's Republic of China
        \and
            CAS Key Laboratory of Optical Astronomy, National Astronomical Observatories, Chinese Academy of Sciences, Beijing 100101, China 
        \and
            School of Astronomy and Space Science, University of the Chinese Academy of Sciences, Beijing 101408, China
        \and
            Department of Astronomy, China West Normal University, Nanchong, China
             }

   \date{Received: \,\,\,\,\,\,\,\,accepted  }

% \abstract{}{}{}{}{} 
% 5 {} token are mandatory
 
  \abstract
   {Surface brightness-color relations (SBCRs) are essential for estimating distances and stellar properties. Previously, SBCRs were based on a limited number of red clump giant (RC) stars, and red giant branch (RGB) stars with infrared interferometric measurements or eclipsing binaries with high-precision radii measurements, resulting in discrepancies in precision and accuracy. Recently, the large number of RGB stars with asteroseismic parameters and precise Gaia distance measurements has enabled the development of more accurate and robust SBCRs.}
   {Asteroseismic radius and Gaia distance (ARD) method has been proposed to establish the SBCRs for late-type stars. }
   {We select Kepler RGB stars with high-precision asteroseismic radii (uncertainties $<$ 1\%) and cross-match them with 2MASS, APASS, and Gaia to obtain Johnson-$B$, Johnson-$V$, $G$, $J$, $H$, and $K_{s}$-band photometric data. After applying selection criteria, we obtain 626 RGB stars to build the SBCR. Among these, 100 RGBs are used as independent validation for the distance, and the remaining samples are used to fit the SBCR.}
  {First, using 526 targets with asteroseismic radii and Gaia distances, 
  nine SBCRs are proposed based on 2MASS ($J$, $H$, $K_s$), APASS (Johnson-$B$, Johnson-$V$), and Gaia ($G$) photometry. The average \texttt{rms} scatter in these relations is 0.075 mag, which corresponds to an uncertainty of approximately 3.5\% in distance. These relations are further validated using 100 independent samples with Gaia distances, showing no bias, with a dispersion of approximately 3\%. Compared to interferometric measurements, a systematic underestimation of 2.3\% was observed, and the discrepancy decreases as the angular diameter increases. Additionally, the distances of eclipsing binaries in the Large Magellanic Cloud and Small Magellanic Cloud obtained using our SBCRs are generally consistent with those measured in the literature, with a dispersion of 1\% and a slight overestimation of 1\% to 2.5\%.}
 {The ARD method capitalizes on two key advantages for precise stellar distance determination: a statistically robust sample of homogeneous RGB stars with low observational costs, and independent distance verification through Gaia data. Such SBCRs can be further calibrated and expanded more efficiently and effectively.}

\titlerunning{SBCR for RGB Stars using Asteroseismic radii and Gaia DR3}
\authorrunning{Jianping Xiong et al.}

\maketitle
% \titlerunning
%

\section{Introduction}
Accurate distance measurement is a cornerstone of astrophysics, underpinning our understanding of the structure, scale, and expansion of the Universe. Determining precise distances to celestial objects enables us to derive fundamental properties such as luminosity and size, which are essential for testing theoretical models and exploring cosmic phenomena \citep{2010ARA&A..48..673F}. Among the various distance determination methods, surface brightness-color relations (SBCRs) have emerged as powerful tools for estimating stellar angular diameters from photometric data \citep{1976MNRAS.174..503B, 1999PASP..111.1515V, 2004A&A...426..297K,2005MNRAS.357..174D}, thereby facilitating accurate distance measurements to stars and their host galaxies  \citep{1969MNRAS.144..297W, 1977ApJ...213..458L, 2013Natur.495...76P, 2019Natur.567..200P, 2020ApJ...904...13G}. With the angular diameters derived from SBCR, when the distance is known, applying the SBCR can directly yield stellar radius. Conversely, if the radius is known, a reliable distance can be calculated.

As a successful application, \citet{2019Natur.567..200P} calibrated the SBCR using 41 nearby red clump giant stars (RCs) within the $V-K$ color range of 2.1 to 2.8. These RCs were observed with precise near-infrared photometry at the South African Astronomical Observatory \citep{2012MNRAS.419.1637L}, and their angular diameters were measured using the ESO VLTI and PIONIER instruments \citep{2018A&A...616A..68G} with a precision of 1\%. Combining with 20 eclipsing binaries with radii determination, this relation was further used to measure the distance to the Large Magellanic Cloud (LMC) with a precision of 1\%. Additionally, the same SBCR was employed to estimate the distance to the Small Magellanic Cloud (SMC), achieving a precision of 2\% \citep{2020ApJ...904...13G}.

%The surface brightness-color relation also provides a convenient method for expressing surface brightness as a function of intrinsic stellar color. It holds significant scientific importance in astrophysics by offering precise techniques for measuring stellar angular diameters \citep{1976MNRAS.174..503B, 1999PASP..111.1515V, 2004A&A...426..297K,2005MNRAS.357..174D} and plays a crucial role in determining extragalactic distances \citep{1977ApJ...213..458L, 2013Natur.495...76P, 2019Natur.567..200P, 2020ApJ...904...13G}. When the distance is known, applying the SBCR can directly yield stellar radius. Conversely, if the radius is known, a reliable distance can be calculated.

To calibrate the SBCRs, the angular diameters obtained through long-baseline interferometry (LBI) are usually required. Besides the SBCR presented by \citet{2019Natur.567..200P}, \citet{2020A&A...639A..67N} calibrated the SBCR for red giant stars in the 2.1 $\leq$ $V-K$ $\leq$ 2.5 color range using 8 G/K-type giants with the homogeneous VEGA/CHARA interferometric data. It demonstrated that the average precision on the limb-darkened angular diameters was 2.4\%. Moreover, \citet{2020A&A...640A...2S, 2021A&A...652A..26S} collected approximately 70 F5/K7 type giants with measured angular diameters from the JMMC Measured Stellar Diameters Catalog (JMDC) \citep{2016yCat.2345....0D} to develop the SBCR within the $V-K$ color range of 1.8 to 3.8. Additionally, they also established SBCRs for F5/K7 type dwarfs and sub-giants, M dwarfs and sub-giants, and M giants. The precision of the angular diameter measurements was estimated to be approximately 1\% to 2.7\%.

The advent of space missions such as Gaia has provided unprecedented parallax data, the SBCRs were also established by detached eclipsing binaries. As we all know, accurate stellar radii can be determined through the combination of time-domain spectroscopic and photometric observations of eclipsing binaries \citep{2023AJ....165...30X, 2010A&ARv..18...67T}, in conjunction with the distance (or trigonometric parallax), thus facilitating the calibration of the SBCRs. For example, \citet{2021A&A...649A.109G} used 28 eclipsing binaries with accurately determined radii and Gaia EDR3 parallaxes to construct the SBCRs, except for $V-K$ color, they also established the SBCRs combined the Johnson $B$, Gaia $G$ and 2MASS $JHK_{s}$ bands. However, these relations only cover dwarf and subgiant stars. The precision of angular diameters prediction for A-, F-, and G-type dwarf and subgiant stars was $\sim$ 1\%.

Recent advancements aim to address these challenges by expanding SBCR calibrations to larger and more diverse samples of stars with high-precision measurements. To date, SBCRs have been proposed for various spectral types, covering both dwarfs and giants. And more than 20 distinct SBCRs have been developed \citep{2020A&A...640A...2S, 2024A&A...690A.327V}, most of which are consistent within the $V-K$ range of 1.5 to 2.5 \citep{2023A&A...671A..14N}. Beyond this range, these SBCRs exhibit significant discrepancies. The precision and robustness of the SBCR are highly dependent on the quality and quantity of the samples. However, the observational costs of long-baseline interferometry are relatively high, and there is a scarcity of detached eclipsing binary samples with high-precision time-domain spectroscopic and photometric data. Consequently, the establishment of SBCRs has been limited by the small number of available samples. Even among giants of the same spectral type, their $V-K$ color ranges are often truncated. This limitation adversely affects the uncertainty of SBCRs.

Fortunately, thanks to the high-precision time-domain photometric data provided by the space missions Kepler \citep{2010Sci...327..977B} and Transiting Exoplanet Survey Satellite (TESS, \citet{2015JATIS...1a4003R}), numerous studies have produced large samples (exceeding several thousand) of RGB and RC stars with parameters such as stellar radii, ages, and masses determined using high-precision asteroseismic methods \citep{2022ApJ...927..167L, 2023ApJ...953..182W, 2024AJ....167...50S}. The typical precision of radius is 4\% \citep{2014ApJS..215...19P, 2015MNRAS.451.2230M, 2024A&A...685A.150V}. Among these, \citet{2023ApJ...953..182W} presented a catalog of 1,153 Kepler red giant branch stars, with the radii, masses and ages of these stars having been determined using asteroseismic model. They combined radial-mode oscillation frequencies, gravity-mode period spacings, Gaia luminosities, and spectroscopic data from LAMOST ((Large Sky Area Multi-Object fiber Spectroscopic Telescope, \citet{2012RAA....12.1197C, 2020arXiv200507210L}) and APOGEE (Apache Point Observatory Galactic Evolution Experiment, \citet{2017AJ....154...94M}) to characterize these stars, and the precision of radius measurements reached around 1\%. 
Moreover, Gaia DR3 observed nearly 1.5 billion targets, the high-precision trigonometric parallaxes and distances are provided \citep{2021AJ....161..147B, 2023A&A...674A...1G}. This extensive data will significantly enhance our calibration of SBCRs for RGB stars.

Therefore, in this paper, we aim to establish an empirical surface brightness-color relation for red giant branch stars based on their asteroseismic radii and Gaia DR3 distances. The structure of the paper is as follows: Section~\ref{data} describes the sample selection and the reduction of observational data. In Section~\ref{method}, we present the establishment of the SBCR. Section~\ref{result} presents the validation of the SBCR in this paper. Finally, Section~\ref{summary} discusses the results and provides a summary of our findings.

%__________________________________________________________________

\section{Data} \label{data}
In this paper, we obtain our initial samples from \citet{2023ApJ...953..182W}. We first compile a sample of 1,153 RGB stars, each with a radius measurement precision of 1\%. Next, we acquire their distances and multi-band photometry. In this section, we detail the selection criteria used to refine our samples.

\subsection{Gaia distance}
Gaia DR3 has provided 1.5 billion stars with parallax measurements, However, due to the nonlinear inverse relationship between parallax and distance and the significant measurement uncertainties (especially for distant stars), directly converting parallax values into distances can result in large errors and unphysical negative distances. \citet{2021AJ....161..147B} transferred the parallaxes to distances using a Bayesian approach based on a prior Galactic spatial distribution. We cross-match the initial sample of 1,153 stars with the geometric distances from \citet{2021AJ....161..147B} and obtain G-band photometry by cross-matching with Gaia DR3 data. After cross-matching, we exclude samples with \texttt{RUWE} $>$ 1.4 and those with distance uncertainties exceeding 10\%. Our estimates of distance uncertainties ($\sigma_{d}$) are as follows:
\begin{equation}\label{sigmad}
    \sigma_{d} = \frac{1}{2}({d_{upper}}-{d_{lower}})
\end{equation}
where $d_{lower}$ and $d_{upper}$ are the 16$\rm{th}$ and 84$\rm{th}$ percentiles of the distances presented by \citet{2021AJ....161..147B}. %Fig.\ref{fig:dist_err} shows the distribution of the distance uncertainties of our samples.

\subsection{Photometric selection}
\subsubsection{Visible photometry}
The American Association of Variable Star Observers (AAVSO) Photometric All-Sky Survey (APASS) project aims to bridge the observational gap left by the shallow two-bandpass Tycho-2 photometric catalog. APASS covers a magnitude range from 10.5 \texttt{mag} to \texttt{mag}. In APASS DR9\footnote{\url{https://www.astroplanner.net/apass.html}}, the project has provided Johnson-$B$, $V$ and Sloan $g$, $r$, $i$ band observations for a total of 61,176,401 stars, achieving a photometric accuracy of 0.02 \texttt{mag} \citep{2015AAS...22533616H, 2016yCat.2336....0H}. In Gaia DR3, magnitudes in the $G$, $BP$, and $RP$ bands are provided for 1.5 billion stars. The high-precision magnitudes in the $G$ band are measured through the Image Parameter Determination (IPD) process, which employs a complex model that incorporates extensive calibrations and a shape-based Point Spread Function (PSF). In contrast, magnitudes in the $BP$ and $RP$ bands are extracted from low-resolution spectra \citep{2021A&A...649A...3R}. Therefore, in this paper, we only include the Gaia $G$-band magnitudes.

\subsubsection{Infrared photometry}
For the infrared magnitudes, obtaining accurate infrared $K$ magnitudes is a challenge, therefore, in previous studies, such as \citet{2020A&A...640A...2S}, both $K$ and $K_{\rm{s}}$ photometries were utilized to calibrate SBCRs. These examinations demonstrated that SBCRs can reliably incorporate both 2MASS $K$ and Johnson $K$ photometries without introducing significant biases (within 2\%) in angular diameter measurements. Currently, the Two Micron All Sky Survey (2MASS) remains the most comprehensive all-sky infrared photometric survey, conducting near-infrared observations in the $J$ (1.25 $\mu m$), $H$ (1.65 $\mu m$), and $K_{\rm{s}}$ (2.16 $\mu m$) bands. The 2MASS Point Source Catalog  (PSC) \citep{2006AJ....131.1163S} includes 471 million stars. This can provide us with a complete set of infrared magnitude measurements. Consequently, in this paper, we employ the infrared magnitudes in the $K_{\rm{s}}$ band of 2MASS to construct the SBCRs.

To ensure the reliability of our data, we included only photometric measurements with errors below 0.1 \texttt{mag}. Additionally, we performed a 1 $arcsec$ cone search to match magnitudes in the Johnson-$B$, $V$, $J$, $H$, and $K_{\rm{s}}$ bands from APASS DR9 and 2MASS. For selected RGB samples, the mean measurement errors are 0.027 \texttt{mag} and 0.026 \texttt{mag} in the Johnson-$B$ and $V$ bands, 0.022 \texttt{mag}, 0.021 \texttt{mag}, and 0.017 \texttt{mag} in the $J$, $H$, and $K_{\rm{s}}$ bands, respectively, and 0.003 \texttt{mag} in the $G$ band.

\section{Establishment of SBCRs} \label{method}
\subsection{Surface brightness calculation}
The surface brightness of a star is defined as the flux density emitted per unit angular area. It is related to its intrinsic reddening-free visual magnitude and its limb-darkened angular diameter. This surface brightness can be calculated using the equation presented in \citet{1969MNRAS.144..297W}, as the following shows:

\begin{equation}
    S_{\lambda} = m_{\lambda0} + 5 log \theta_{LD} \label{Surface}
\end{equation}

In Eq.\ref{Surface}, $S_{\lambda}$ represents the surface brightness in a specific band, in this paper, $\lambda = (B, V$, and $G)$. $m_{\lambda0}$ is the reddening-free visual magnitude in $B$, $V$, and $G$ bands. $\theta_{LD}$ is the limb-darkened angular diameter measured from infrared interferometry. In this paper, $\theta_{LD}$ is calculated by:

\begin{equation} \label{theta}
    \theta_{LD}=\frac{2R}{d} 
\end{equation}

where $R$ is the asteroseismic radius of a star, $d$ is the distance obtained from Gaia provided by \citet{2021AJ....161..147B}.

To obtain the reddening-free visual magnitude, we measure the color excess $E(B-V)$ using the 3-D dust map provided by \textit{dustmaps bayestar 2019} \citep{2019ApJ...887...93G}. During this process, distance is used as a prior to calculate $E(B-V)$. The visual extinction $A_{\rm V}$ is then derived using the total-to-selective extinction ratio $R_{V} = 3.1$. Extinctions in other bands are subsequently determined using the extinction coefficients from Table 3 in \cite{2019ApJ...877..116W}, specifically: $A_{B}$ = 1.317 $A_{V}$, $A_{G}$ = 0.789 $A_{V}$, $A_{J}$ = 0.243 $A_{V}$, $A_{H}$ = 0.131 $A_{V}$, $A_{K_{s}}$ = 0.078 $A_{V}$.

We then calculate the surface brightness by combining the asteroseismic radii with Gaia distances using Eq.~\ref{Surface} and Eq.~\ref{theta}. The Gaussian distributions are constructed based on the magnitudes, distances, radii, and extinction values, incorporating their respective uncertainties. To propagate these uncertainties, we perform 2,000 Monte Carlo simulations using these Gaussian distributions. This approach allows us to calculate the surface brightness in the $B$, $V$, and $G$ bands, as well as the extinction-free magnitudes for each band. The uncertainties in these parameters are quantified by the standard deviations of the resulting distributions, which are derived from the Monte Carlo simulations. Finally, we exclude samples with surface brightness errors exceeding 0.1 \texttt{mag}, resulting in a total of 626 RGB samples. The mean errors of surface brightness in the $B$, $V$, and $G$ bands are $\sigma_{S_{B}}$ = 0.055 \texttt{mag}, $\sigma_{S_{V}}$ = 0.056 \texttt{mag}  and $\sigma_{S_{G}}$ = 0.044 \texttt{mag} respectively. We randomly select 100 RGB samples for internal validation of the SBCRs and used the remaining 526 RGB samples to fit the SBCRs.

\begin{figure}[h]
    \centering
    \includegraphics[width=0.45\textwidth]{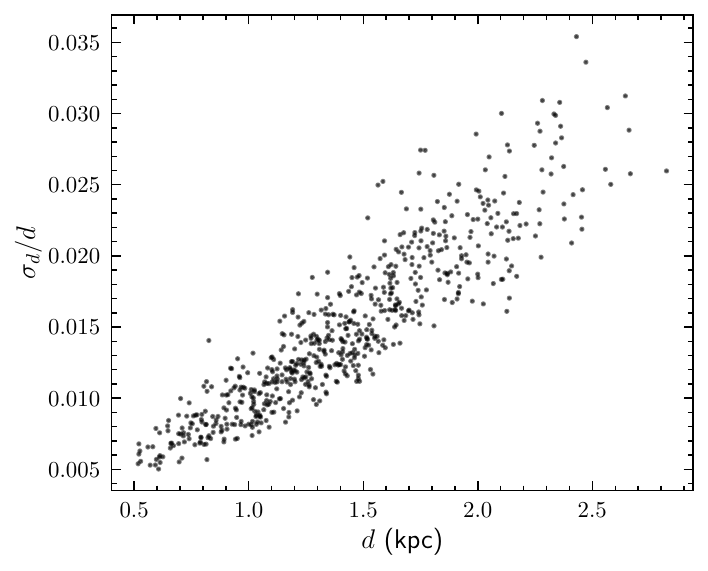}
    \caption{Relationship between distance and distance error for our sample. The distance uncertainties ($\sigma_{d}$) are calculated by Eq.\ref{sigmad}. The average distance error of 1.5\%.} \label{fig:distance_distribution}
\end{figure}

Fig.\ref{fig:distance_distribution} illustrates the distribution of distances and their associated uncertainties ($\sigma_{d}$, which is estimated by Eq.\ref{sigmad}), indicating that distance uncertainties increase with larger distances. It reveals that the selected samples cover a distance range of 0.5 \texttt{kpc} to 2.8 \texttt{kpc}, with the average distance error of 1.5\%.

\subsection{SBCRs fitting}
In this paper, we construct SBCRs in the Johnson-$B$ and $V$ bands and the Gaia $G$ band for RGB stars. We fit these relations using a linear model, $y = ax + b$. Our fitting procedure employs the Orthogonal Distance Regression (ODR) fitting (\textit{scipy.odr}). The ODR method is selected because it simultaneously accounts for uncertainties in both the independent ($x$) and dependent ($y$) variables. This approach minimizes the overall perpendicular distances from the data points to the fitting line, resulting in a more accurate and reliable regression curve compared to traditional least squares methods that only consider errors in the dependent variable. 
\begin{figure}[h]
  \centering
   \includegraphics[scale=0.65]{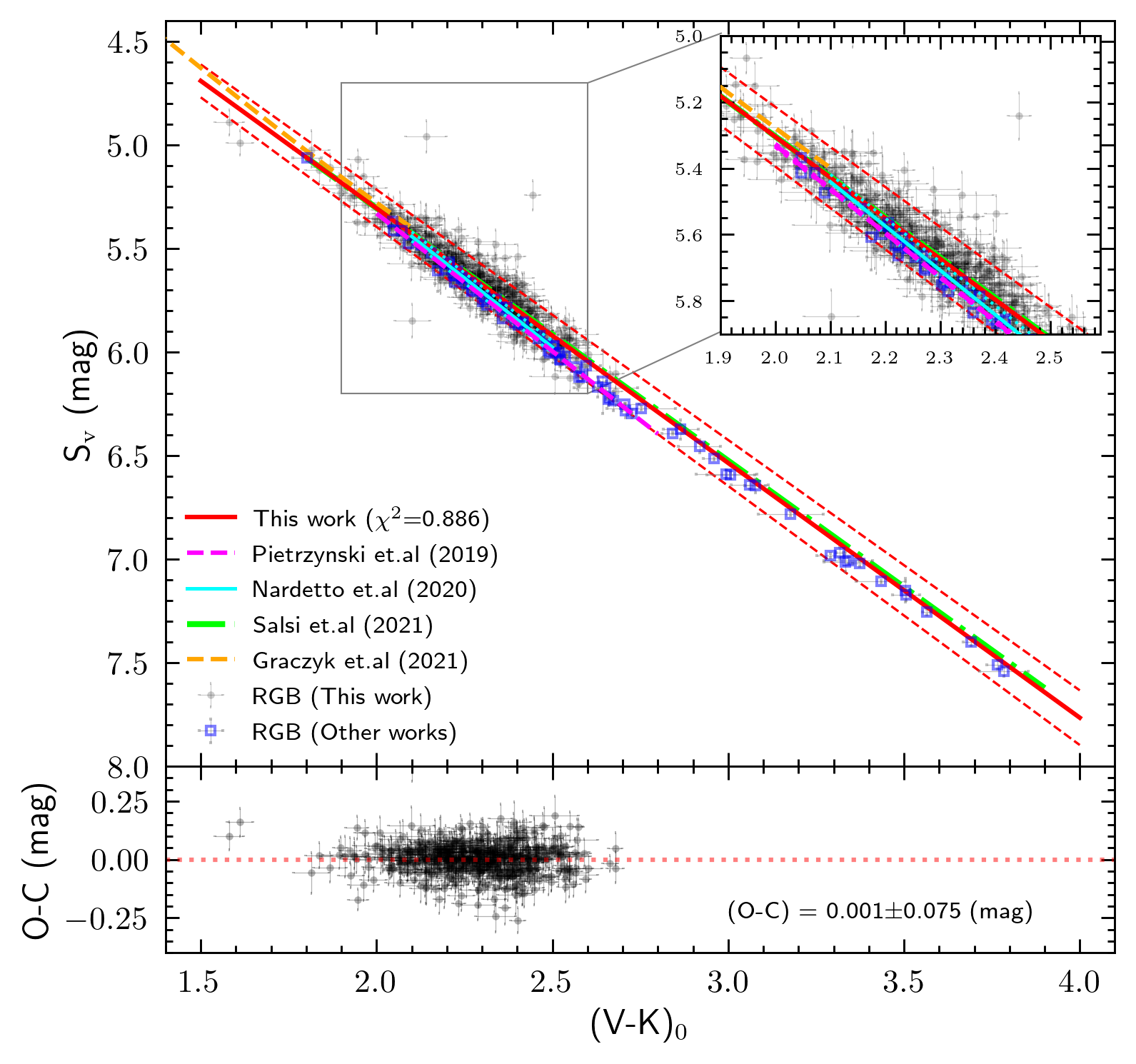}
  \centering
  \caption{Surface brightness-color relations in $V-K$. Black points represent the 526 RGB stars used in this study to fit the SBCR, with their surface brightness estimates derived from asteroseismic radii and Gaia distances. Blue open squares denote the RGB and RC samples from the literature \citep{2019Natur.567..200P, 2020A&A...640A...2S, 2021A&A...652A..26S} used to calibrate the SBCR, with their surface brightness estimates based on angular diameters measured via infrared interferometry. The red solid line illustrates the SBCR fitted in this work, with dashed lines indicating the associated uncertainties ($1\sigma$). The magenta dashed line, the sky blue solid line, the green dotted line, and the orange dashed line represent SBCRs proposed by \citet{2019Natur.567..200P}, \citet{2020A&A...639A..67N}, \citet{2021A&A...652A..26S}, and \citet{2021A&A...649A.109G}, respectively. The lower panel displays the residuals of the fit.} \label{fig:VK_SBCR_compare}
\end{figure}

\begin{figure*}[h]
    \centering
    \begin{subfigure}{0.3\textwidth}
        \includegraphics[width=\textwidth]{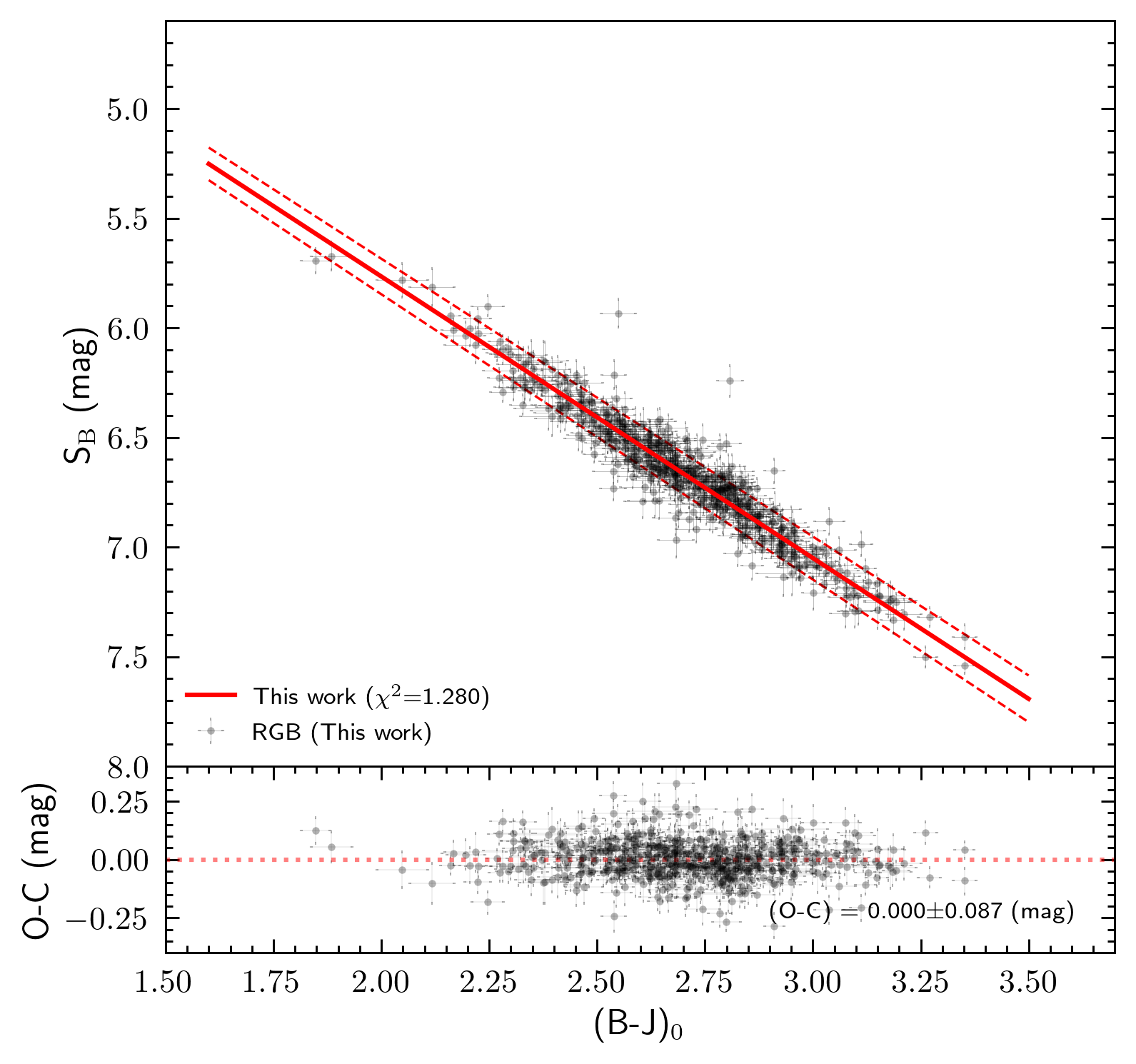}
    \end{subfigure}
    \begin{subfigure}{0.3\textwidth}
        \includegraphics[width=\textwidth]{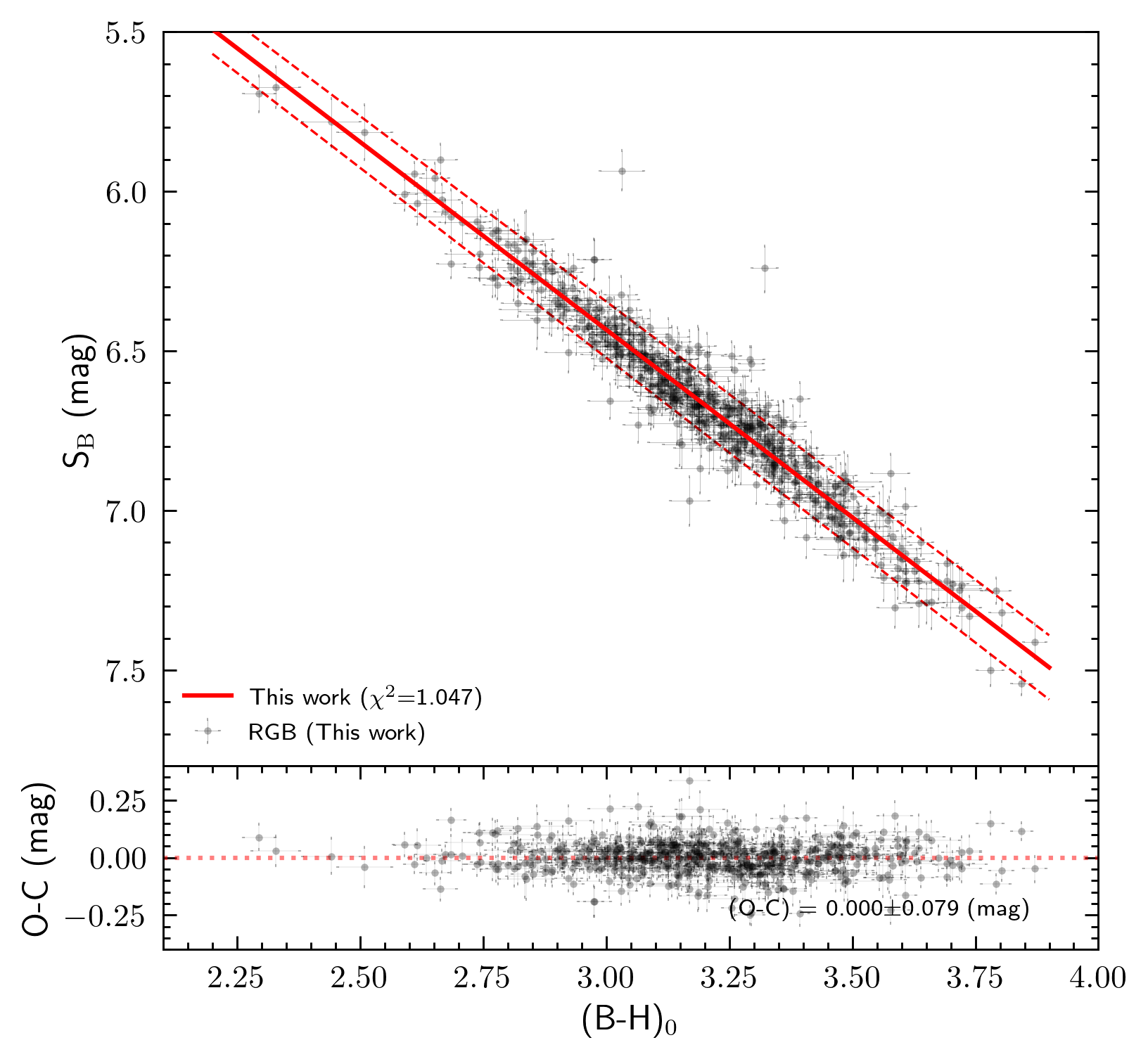}   
    \end{subfigure}
    \begin{subfigure}{0.3\textwidth}
        \includegraphics[width=\textwidth]{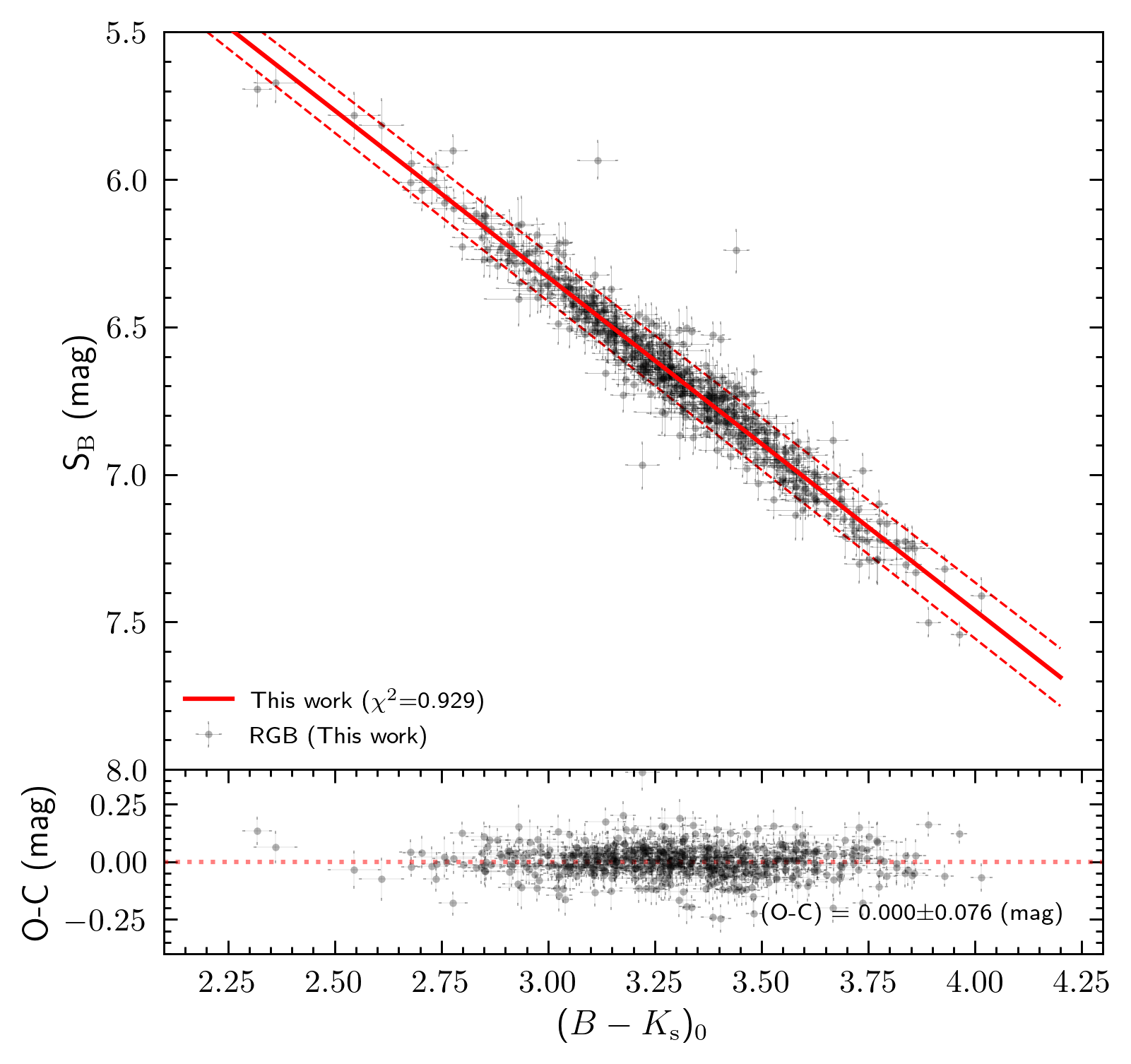}
    \end{subfigure}

    \begin{subfigure}{0.3\textwidth}
        \includegraphics[width=\textwidth]{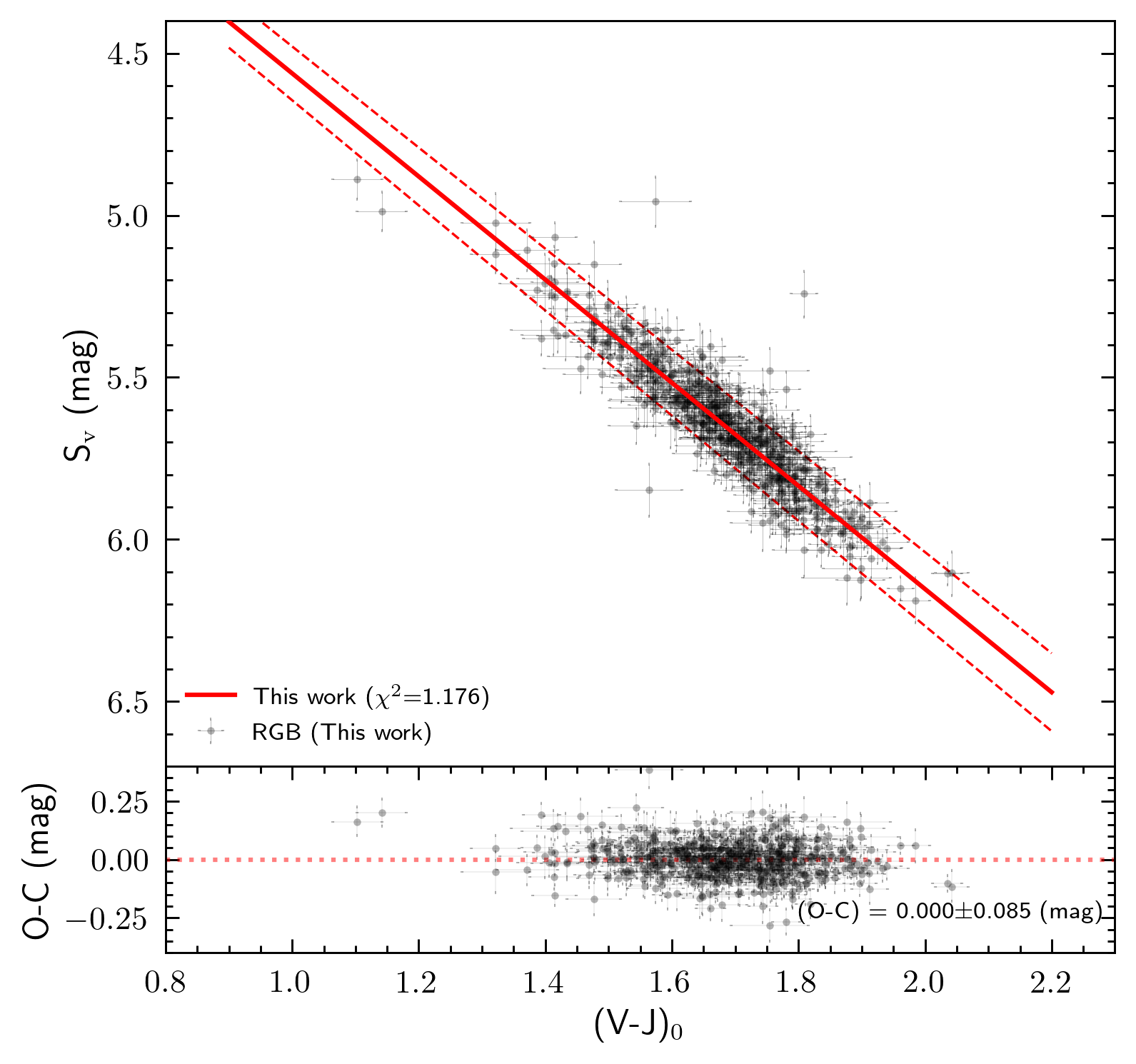}
    \end{subfigure}
    \begin{subfigure}{0.3\textwidth}
        \includegraphics[width=\textwidth]{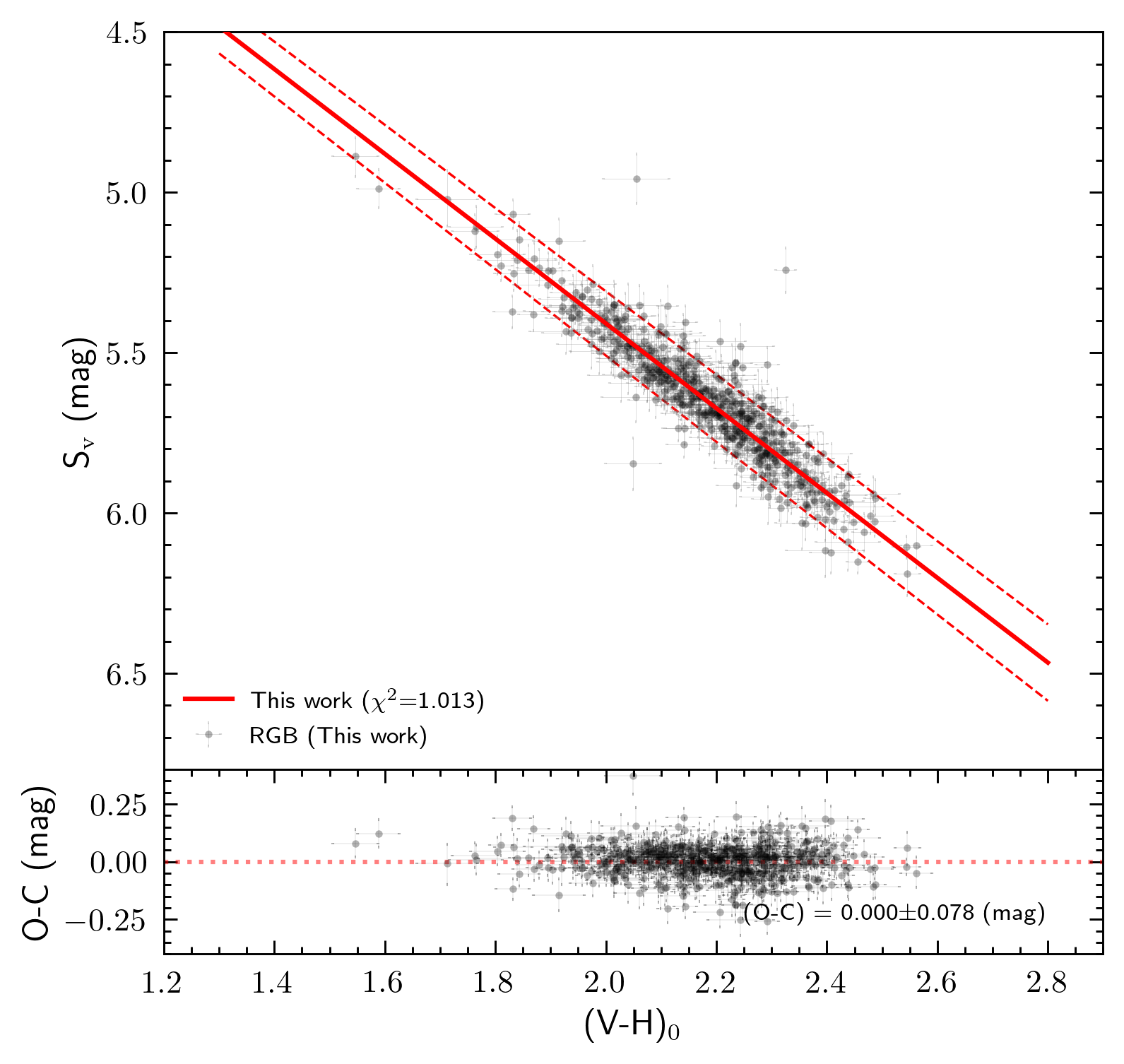}
    \end{subfigure}
    \begin{subfigure}{0.3\textwidth}
        \includegraphics[width=\textwidth]{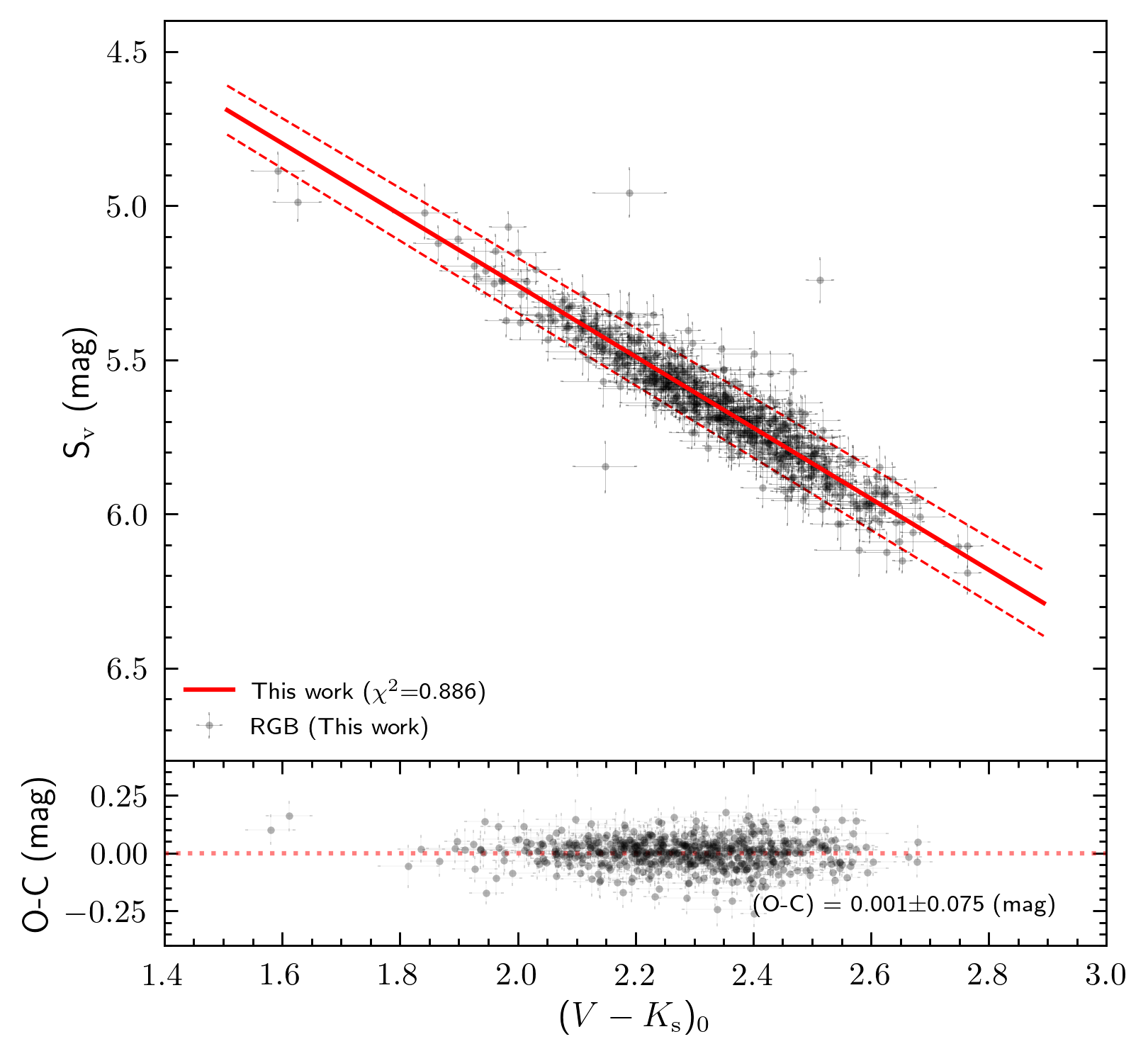}
    \end{subfigure}

    \begin{subfigure}{0.3\textwidth}
        \includegraphics[width=\textwidth]{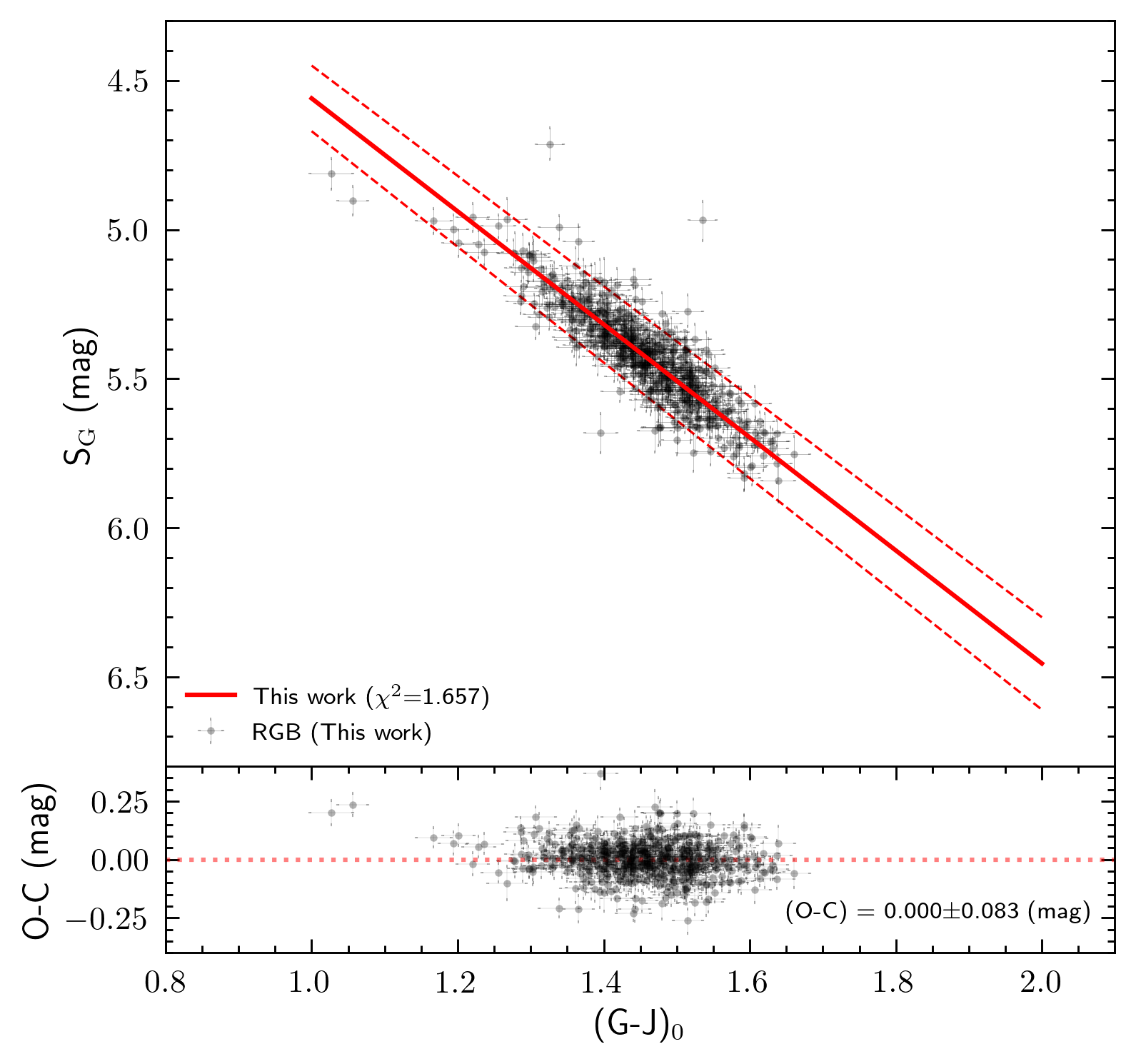}
    \end{subfigure}
    \begin{subfigure}{0.3\textwidth}
        \includegraphics[width=\textwidth]{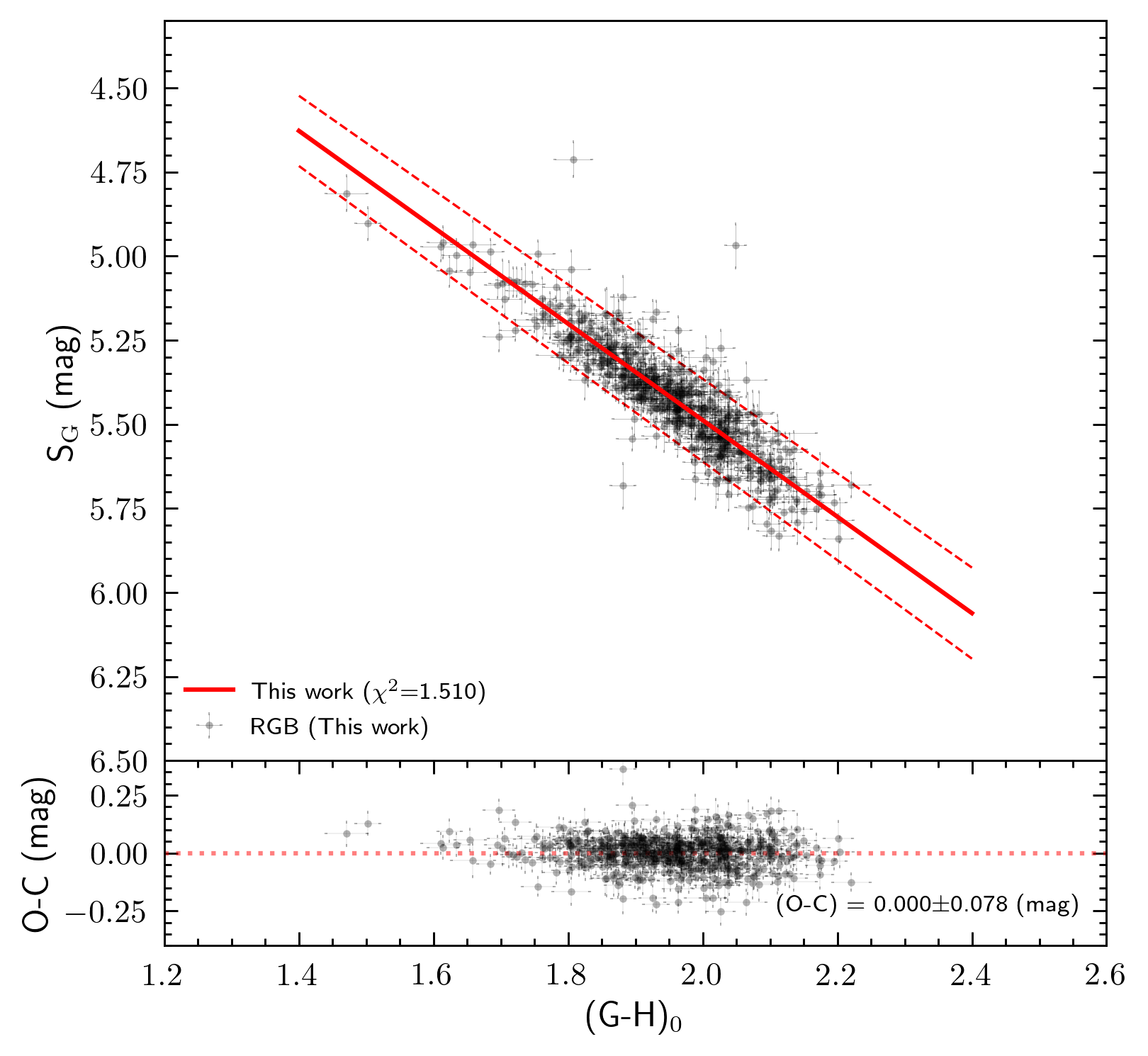}
    \end{subfigure}
    \begin{subfigure}{0.3\textwidth}
        \includegraphics[width=\textwidth]{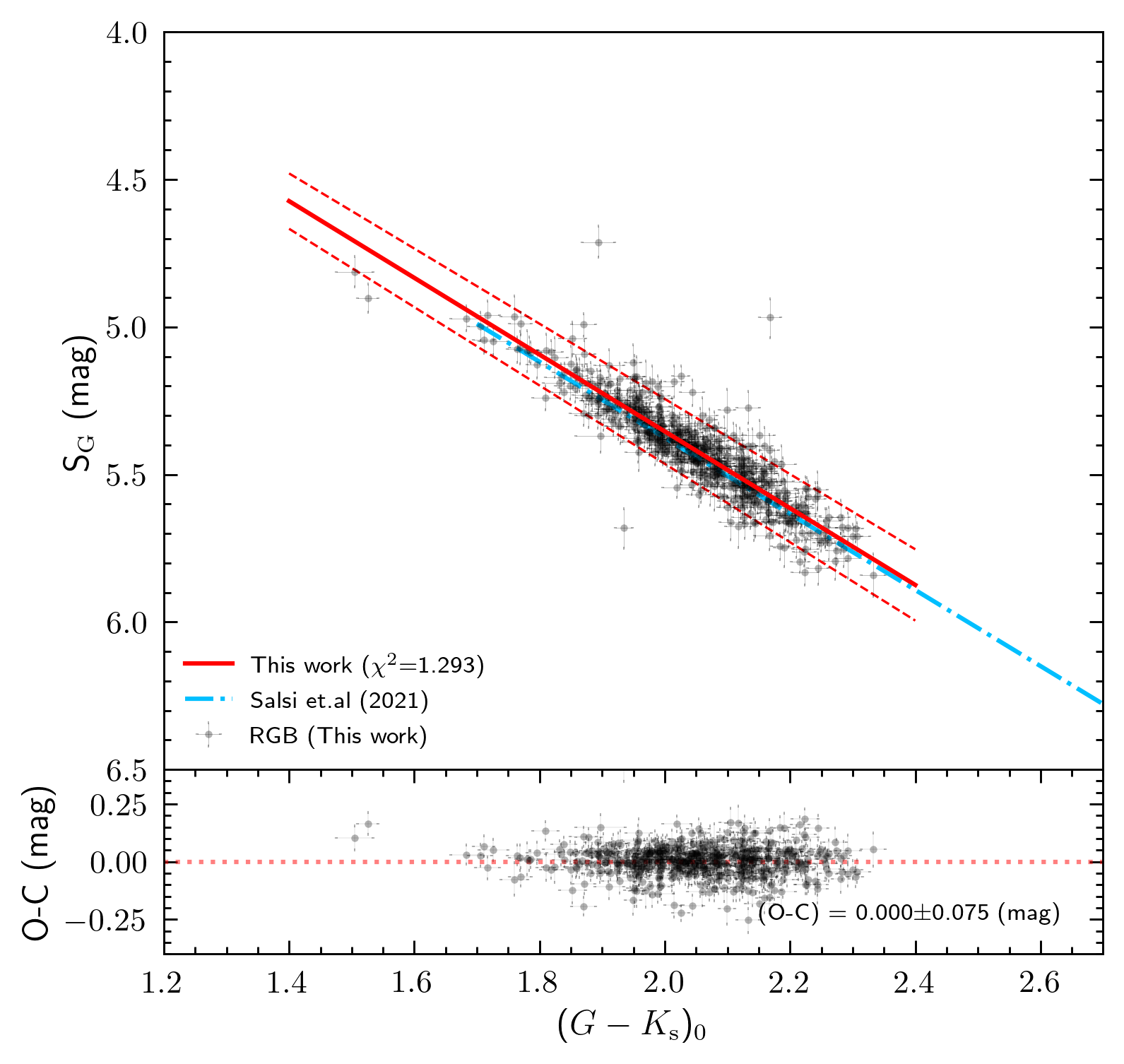}
    \end{subfigure}

    \caption{Fitting results of the surface brightness $S_{\rm{\lambda}}$ in the Johnson $B$ and $V$ Bands and Gaia $G$ band for RGB stars. Black points represent the 526 observed RGB stars. Red solid lines depict the fitting relations, while red dashed lines denote the associated uncertainties. The bottom panel displays the residuals of the fit.}\label{fig:SBCRs_in_9bands}
\end{figure*} 

\begin{figure*}[h]
    \centering
    \begin{subfigure}{0.3\textwidth}
        \includegraphics[width=\textwidth]{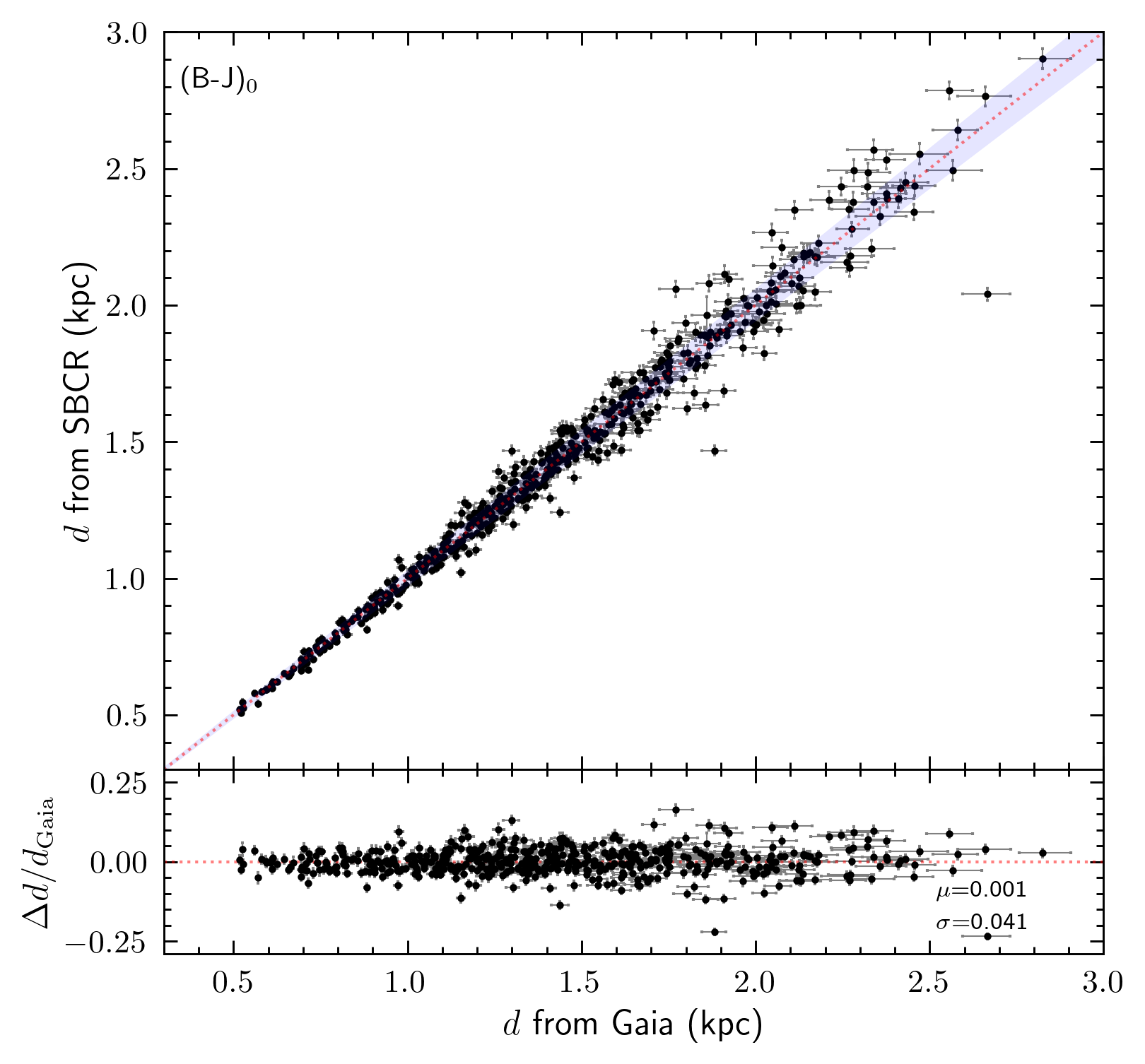}
    \end{subfigure}
    \begin{subfigure}{0.3\textwidth}
        \includegraphics[width=\textwidth]{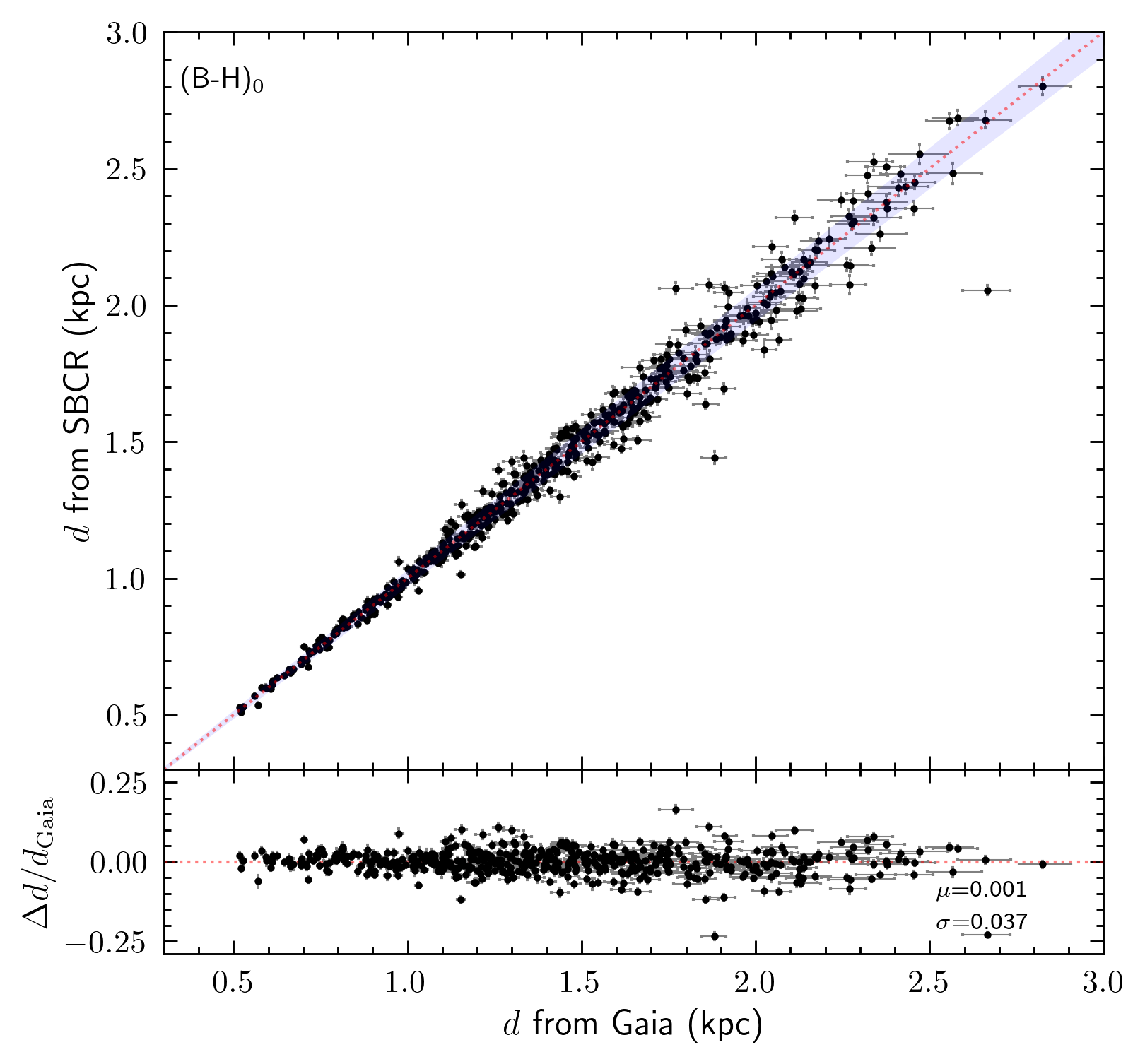}
    \end{subfigure}
    \begin{subfigure}{0.3\textwidth}
        \includegraphics[width=\textwidth]{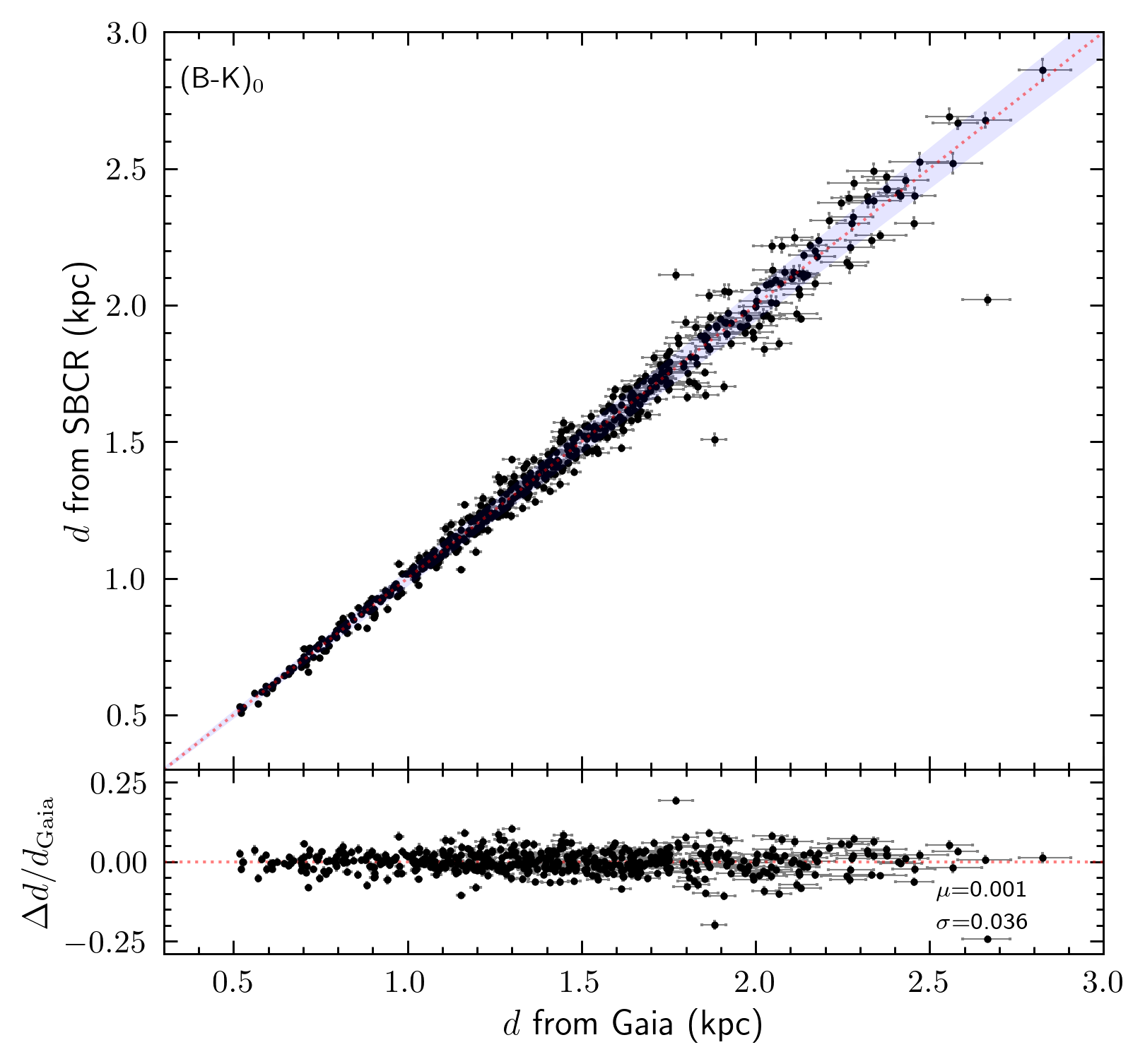}
    \end{subfigure}
    
    \begin{subfigure}{0.3\textwidth}
        \includegraphics[width=\textwidth]{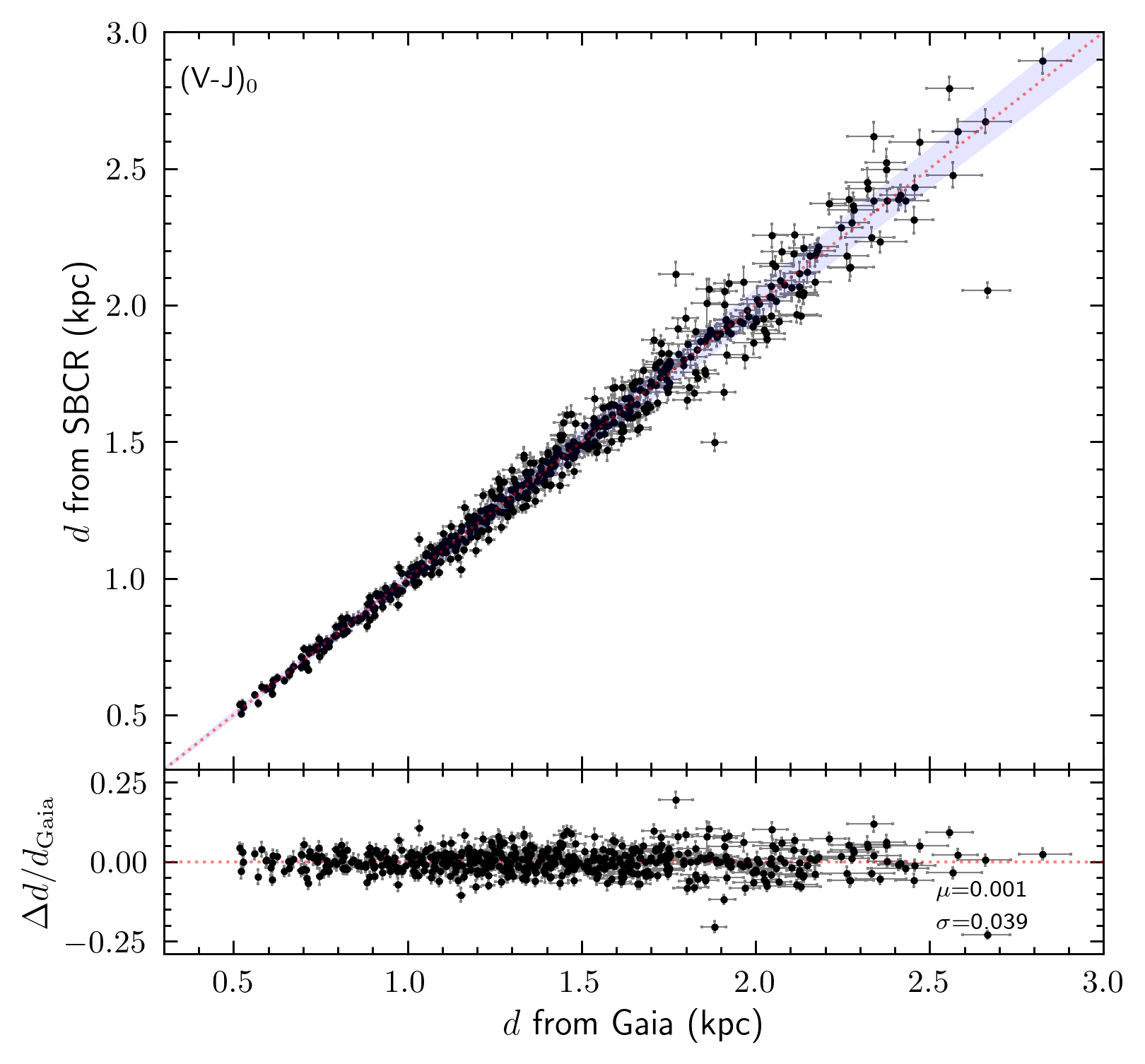}
    \end{subfigure}
    \begin{subfigure}{0.3\textwidth}
        \includegraphics[width=\textwidth]{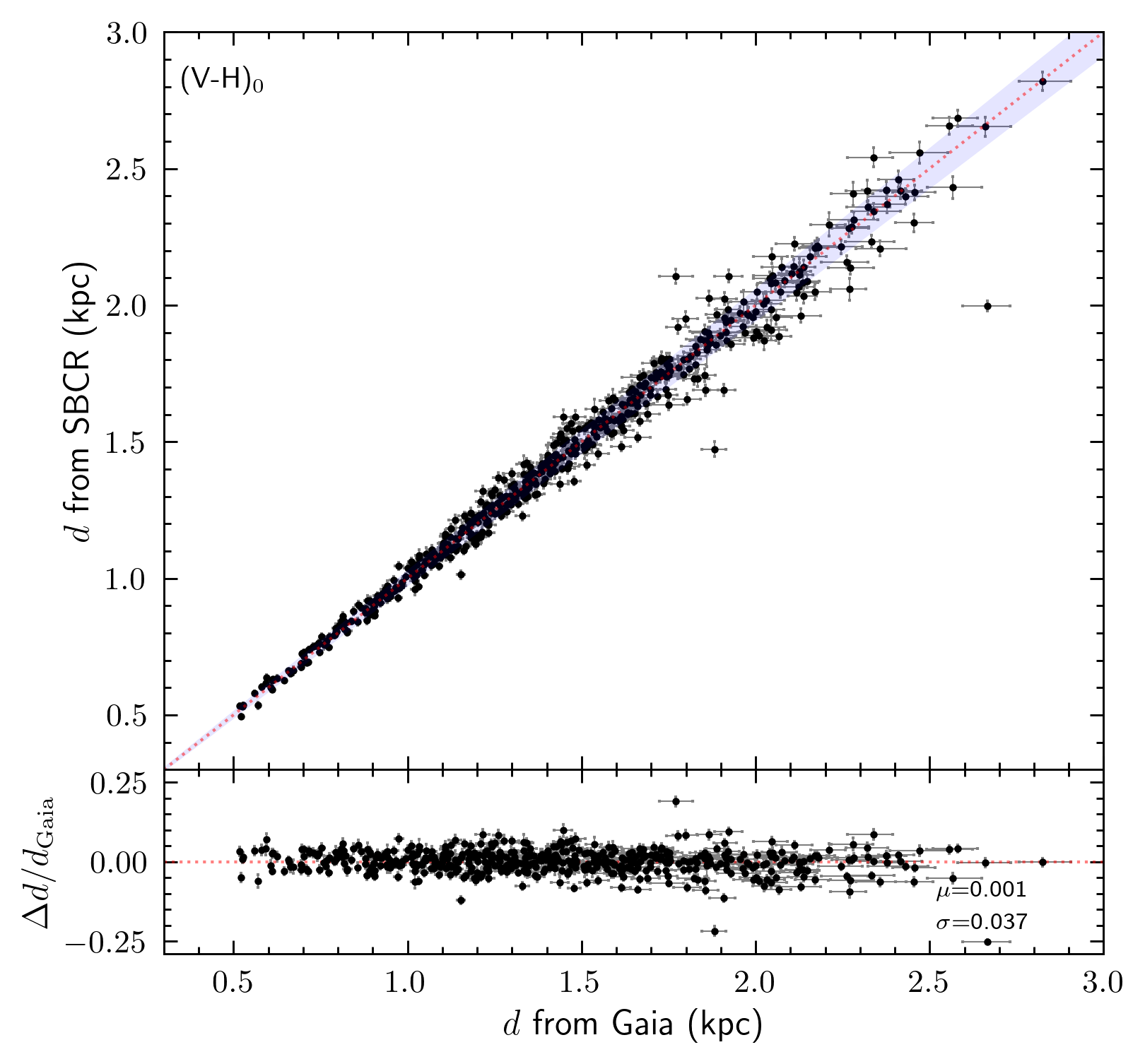}
    \end{subfigure}
    \begin{subfigure}{0.3\textwidth}
        \includegraphics[width=\textwidth]{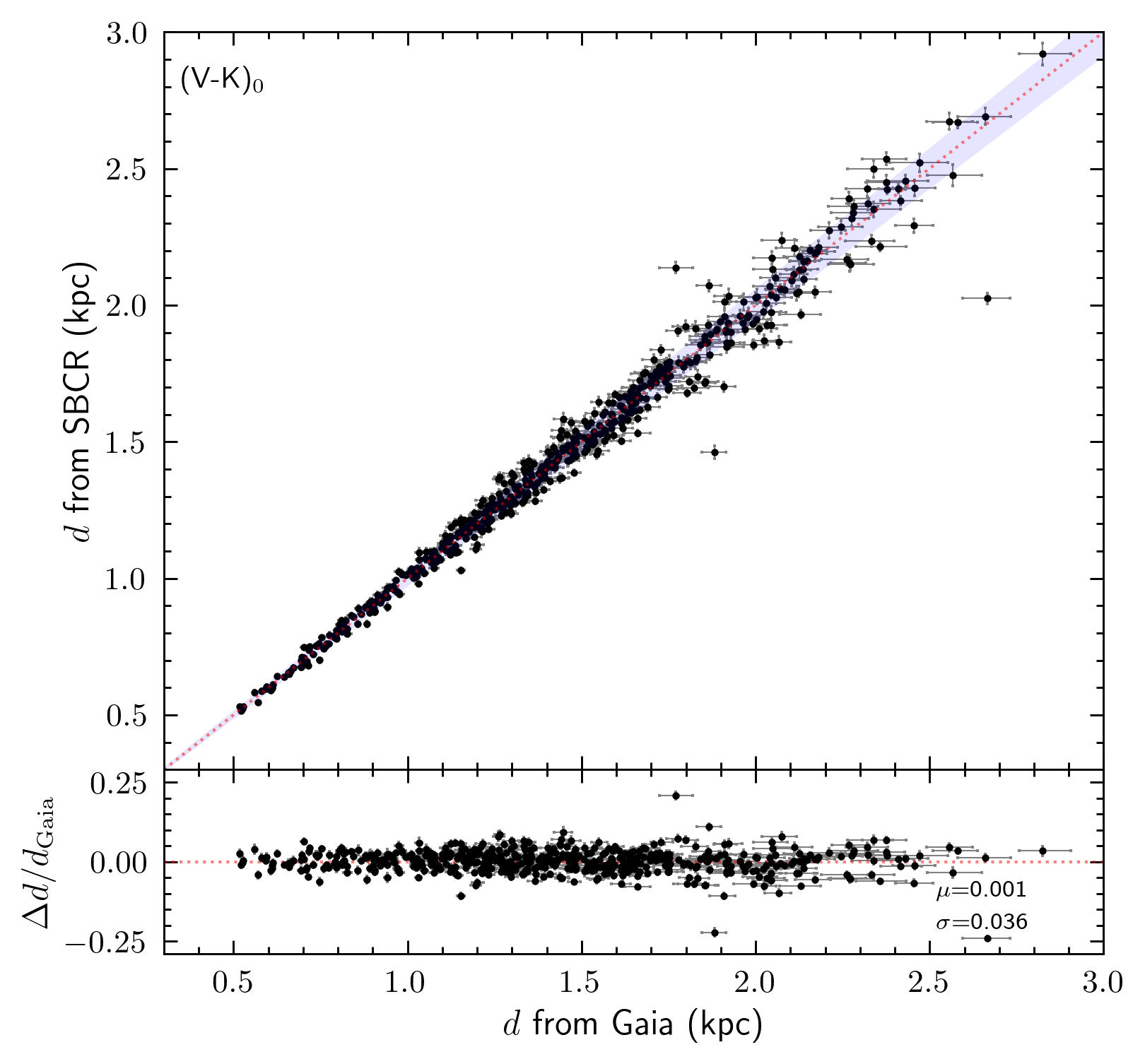}
    \end{subfigure}

    \begin{subfigure}{0.3\textwidth}
        \includegraphics[width=\textwidth]{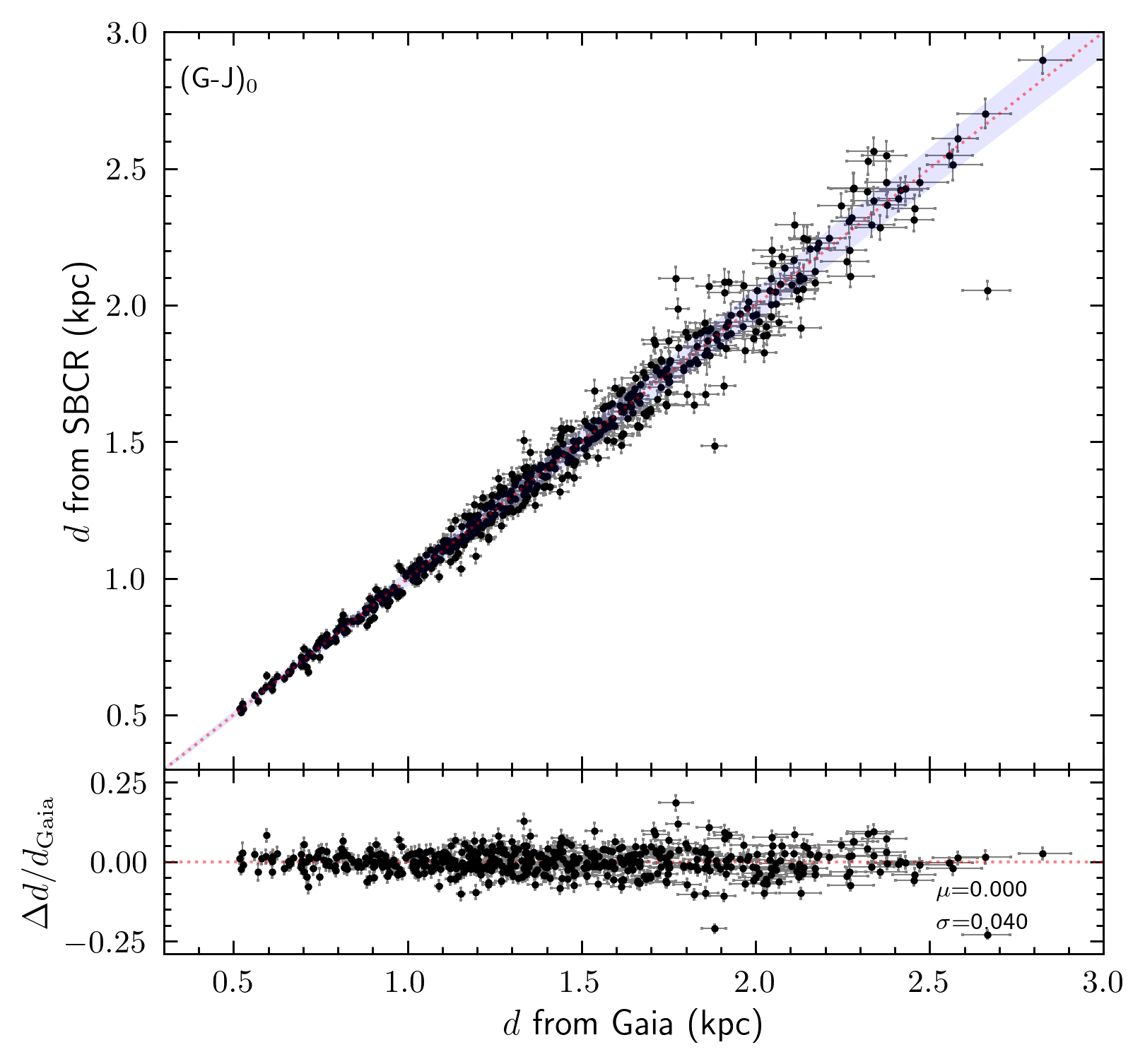}
    \end{subfigure}
    \begin{subfigure}{0.3\textwidth}
        \includegraphics[width=\textwidth]{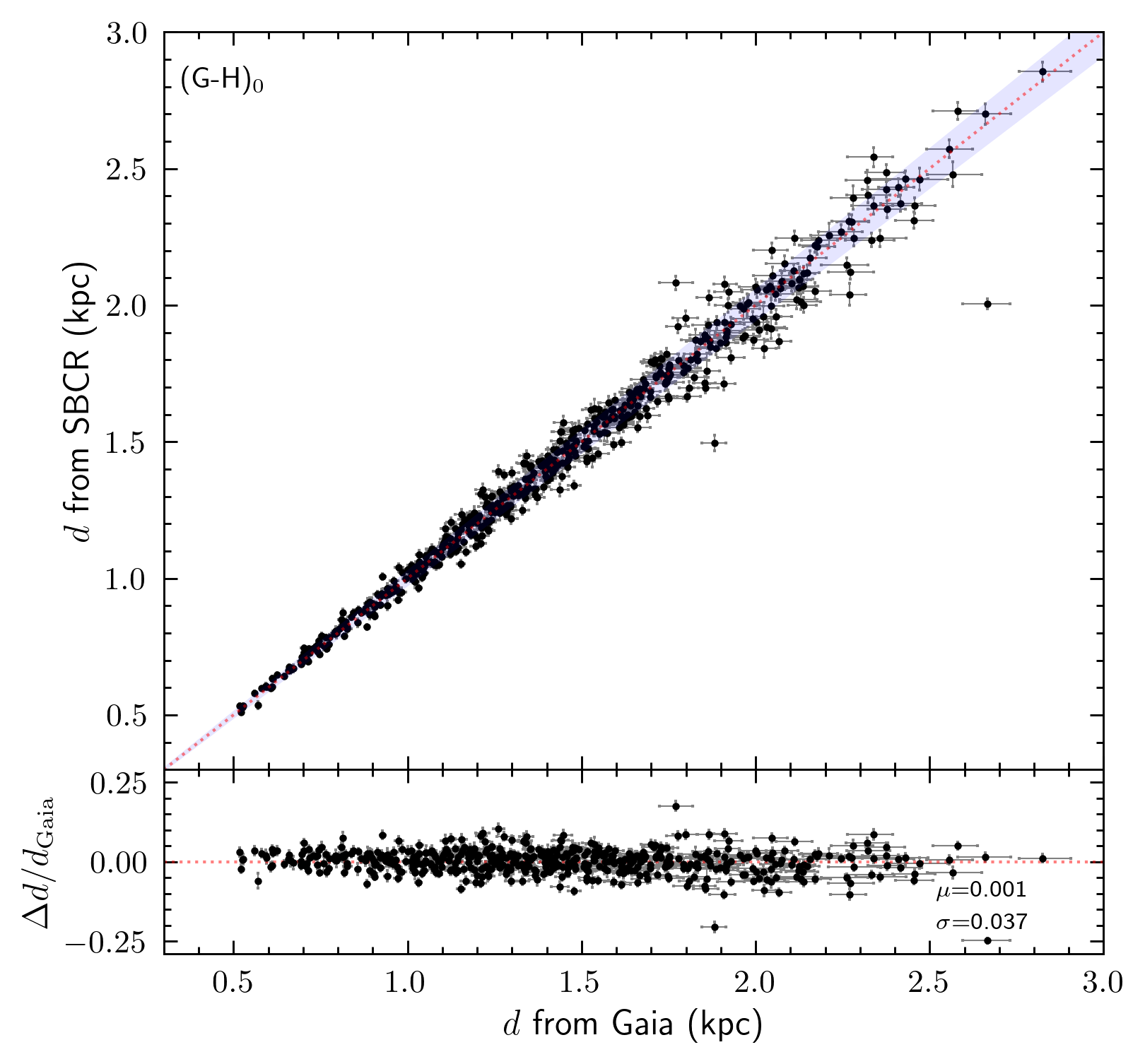}
    \end{subfigure}
    \begin{subfigure}{0.3\textwidth}
        \includegraphics[width=\textwidth]{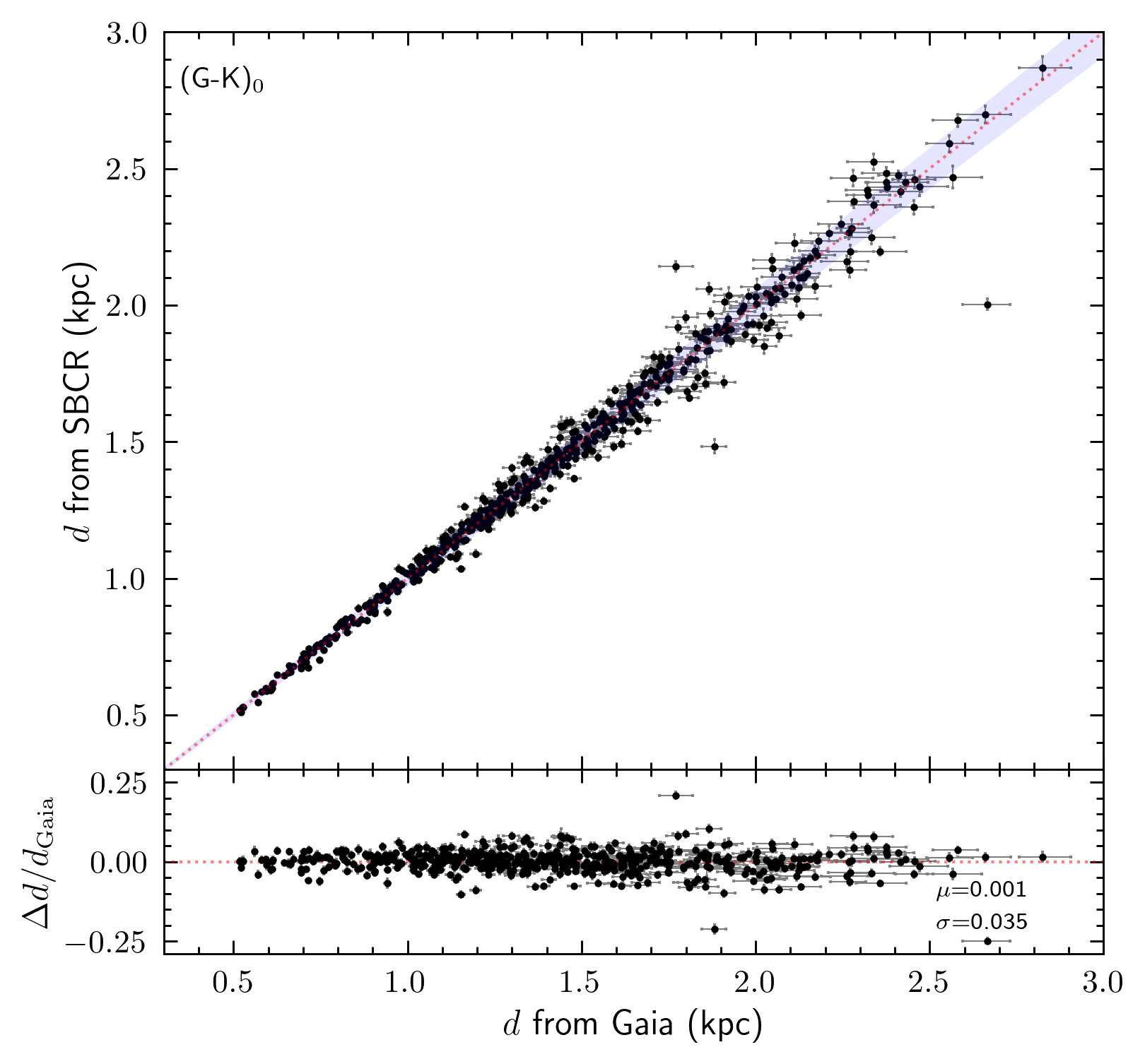}
    \end{subfigure}
    \caption{Validation results of the fitting samples. The $x$-axis represents distances measured by Gaia \citep{2021AJ....161..147B}, while the $y$-axis shows distances estimated using the SBCRs in the Johnson $B$ and $V$ bands and the Gaia $G$ band. The blue shaded area indicates a deviation range of $\pm$3\%. The lower subplot displays the distribution of relative errors ($\Delta d/ d_{Gaia}$, $\Delta d$ is calculated by $d_{\rm{SBCR}}$-$d_{\rm{Gaia}}$).}\label{fig:validation_fitting}
    
\end{figure*}

\begin{table*}[h]
\caption{The fitting coefficients ($a$ and $b$)$^{\rm 1}$ of the surface-brightness $S_{\rm{\lambda}}$ in Johnson $B$ and $V$ and Gaia $G$ bands for RGB stars. }\label{tab:table1}
\centering
\small
\begin{tabular}{cccccc}
\hline
\hline
Band & Color & Color range & $a$ & $b$ & \texttt{rms} ($1\sigma$,\texttt{mag})\\
\hline
\hline
$B$ & $(B-J)_{\rm{0}}$ & [1.8:3.4] & $1.284 \pm 0.017$ & $3.198 \pm 0.047$ & $0.087$\\
$B$ & $(B-H)_{\rm{0}}$ & [2.2:3.9] & $1.176 \pm 0.014$ & $2.905 \pm 0.046$ & $0.079$\\
$B$ & $(B-K_{s})_{\rm{0}}$ & [2.3:4.0] & $1.130 \pm 0.013$ & $2.941 \pm 0.043$ & $0.076$\\
$V$ & $(V-J)_{\rm{0}}$ & [1.1:2.1] & $1.591 \pm 0.031$ & $2.971 \pm 0.052$ & $0.085$\\
$V$ & $(V-H)_{\rm{0}}$ & [1.5:2.6] & $1.322 \pm 0.024$ & $2.765 \pm 0.052$ & $0.078$\\
$V$ & $(V-K_{s})_{\rm{0}}$ & [1.6:2.7] & $1.230 \pm 0.021$ & $2.845 \pm 0.048$ & $0.075$\\
$G$ & $(G-J)_{\rm{0}}$ & [1.0:1.7] & $1.895 \pm 0.045$ & $2.665 \pm 0.065$ &$0.083$\\
$G$ & $(G-H)_{\rm{0}}$ & [1.4:2.2] & $1.434 \pm 0.031$ & $2.620 \pm 0.061$ & $0.078$\\
$G$ & $(G-K_{s})_{\rm{0}}$ & [1.5:2.3] & $1.301 \pm 0.027$ & $2.752 \pm 0.056$ &$0.075$\\
\hline
\end{tabular}

\begin{tablenotes}
\footnotesize
\item[i] 
$^{\rm 1}$ The linear form of the fit is expressed as: $y=ax+b$, $y$ represents the surface brightness in various bands, $x$ represents the different colors.
\end{tablenotes}
\centering
\end{table*}

In Fig.~\ref{fig:VK_SBCR_compare}, we compare our SBCR in $V-K$ (($S_{V}=(1.230\pm0.021)(V-K_{s})_{0}+(2.845\pm0.048)$)) with this from the literature. The magenta dashed line, the sky blue solid line, the green dotted line, and the orange dashed line represent SBCRs proposed by \citet{2019Natur.567..200P} ($S_{V}=(1.330\pm0.017)[(V-K)_{0}-2.405]+(5.869\pm0.003)$), \citet{2020A&A...639A..67N} ($S_{V}=(1.338\pm0.160)[(V-K)_{0}-2.405]+(5.849\pm0.027)$), \citet{2021A&A...652A..26S} ($S_{V}=1.22(V-K)_{0}+2.864$), and \citet{2021A&A...649A.109G} ($S_{V}=1.708(V-K)-0.705(V-K)^{2}+0.623(V-K)^3-0.239(V-K)^{4}+0.0313(V-K)^{5}+2.521$) respectively. Black points indicate our fitting samples, while blue open squares represent the RGB and RC samples compiled from the literature \citep{2019Natur.567..200P, 2020A&A...640A...2S, 2021A&A...652A..26S}. The surface brightness of the literature samples was measured using angular diameters obtained from infrared interferometry and high-precision photometric measurements. Consequently, their dispersion is smaller compared to our samples. Moreover, these samples still fall within the range of our data. The red solid line shows the SBCR fitted in this study, and the red dashed lines represent the uncertainties ($1\sigma$) of this fitted curve. The residuals between the surface brightness predicted by the fitted line ($C$) and the observed surface brightness ($O$) are shown in the bottom panel. 

As shown in Fig.~\ref{fig:VK_SBCR_compare}, the \texttt{rms} scatter ($1\sigma$) of SBCR for $V-K$ in this study is unbiased 0.075 \texttt{mag}. The \texttt{rms} scatters for the relation proposed by \citet{2019Natur.567..200P}, \citet{2020A&A...639A..67N}, and \citet{2021A&A...652A..26S} are 0.018 \texttt{mag}, 0.039 \texttt{mag} and 0.022 \texttt{mag} respectively. It can also be seen that our relation generally aligns with those in the literature in the $V-K$ range of $2.0$ to $2.5$. For example, for a specific color such as $V-K = 2.3$, the $S_v$ values obtained using the relations from \citet{2019Natur.567..200P}, \citet{2020A&A...639A..67N}, and \citet{2021A&A...652A..26S} are $5.729 \pm 0.018$ \texttt{mag}, $5.708 \pm 0.039$ \texttt{mag}, and $5.670 \pm 0.022$ \texttt{mag}, respectively. Using our relation, the $S_V$ = $5.674 \pm 0.075$ \texttt{mag}. Furthermore, our relation agrees well with the samples from the literature, even for larger $V-K$ values.

To assess the goodness of fit, we use the normalized chi-square ($\chi^2$), calculated as Eq.~\ref{Eq:chi}. Additionally, the SBCR we derived exhibits good agreement with the high-precision samples provided in the literature.
\begin{equation}
    \chi^2 = \frac{1}{N - p} \sum_{i=1}^{N} \left( \frac{O_i - C_i}{\sigma_i} \right)^2 \label{Eq:chi}
\end{equation} 

In Eq.~\ref{Eq:chi}, $N$ is the number of data points, $p$ represents the number of fitting parameters (with $p = 2$ in this study), $O_i$ is the observed surface brightness and $C_i$ corresponds to the predicted surface brightness. $\sigma_{i}$ denotes the observational errors.

The surface brightness-color relations in the $V-K$ color index, as well as colors $B-J$, $B-H$, $B-K_{s}$, $V-J$, $V-H$, $G-J$, $G-H$, and $G-K_{s}$ are also derived for RGB stars. Their fitting coefficients ($a$, $b$) are presented in Table ~\ref{tab:table1}, the corresponding errors of the fitting parameters and \texttt{rms} scatters are also shown in the Table ~\ref{tab:table1}. An overview of the 9 SBCRs is shown in Fig.~\ref{fig:SBCRs_in_9bands}.
In Fig.~\ref{fig:SBCRs_in_9bands}, the red solid lines represent our fitted SBCRs, and the red dashed lines indicate the uncertainties ($1\sigma$) of the SBCRs. The bottom panels in each subplot display the residuals between the fitted values and the observations. Black points denote our RGB samples. In the $G-K_{s}$ color index, the blue dashed line corresponds to the $G-K_{s}$ SBCR derived by \citet{2021A&A...652A..26S} (From Table 4 for F5/K7-II/III). It can be seen that the $G-K_{s}$ relation obtained by \citet{2021A&A...652A..26S} falls within the uncertainty range of our fit

Fig.~\ref{fig:validation_fitting} presents a comparison of distance predictions calculated by our SBCRs for the fitting samples. In Fig.~\ref{fig:validation_fitting}, the  $x$-axis represents the geometric distances derived by from Gaia trigonometric parallaxes \citep{2021AJ....161..147B}, while the $y$-axis shows the distances calculated using our SBCRs. The bottom panel displays the distribution of relative residuals between the predicted and observed distances, and the relative residuals ($\Delta d/d_{\rm{Gaia}}$) are derived by $(d_{\rm{SBCR}}-d_{\rm{Gaia}})/d_{Gaia}$, where $d_{\rm{SBCR}}$ is the distance calculated by SBCR. As shown in Fig.~\ref{fig:validation_fitting}, the distances predicted using SBCRs align well with Gaia's geometric distances, exhibiting no bias and an average dispersion of approximately 3\% $\sim$ 4\%. Dispersion increases for more distant targets, but for sources within 1 \texttt{kpc}, the distances we obtain are largely consistent with Gaia's measurements. Some outliers with larger discrepancies may be attributed to observational limitations. Compared to the $K_{s}$ band, the dispersion in the $J$ and $H$ bands is slightly larger. The likely reason is the precision of the 2MASS photometry. For $K_{s}$ band the precision is 1\%, and in the $J$ and $H$ bands, the precision are only 2\%. In our fitting, the overall reliability of the SBCR calibration is shown and the results validate the accuracy of our fitting.

\section{Validation of SBCRs} \label{result}
We established the surface brightness-color relations based on APASS Johnson-$B$ and $V$, Gaia $G$, and 2MASS $J$,$H$, $K_{s}$ photometric data. To further assess the accuracy of these relations, we validated them using both internal and external datasets.

\subsection{Internal validation}

\begin{figure*}[h]
    \centering
    \begin{subfigure}{0.3\textwidth}
        \includegraphics[width=\textwidth]{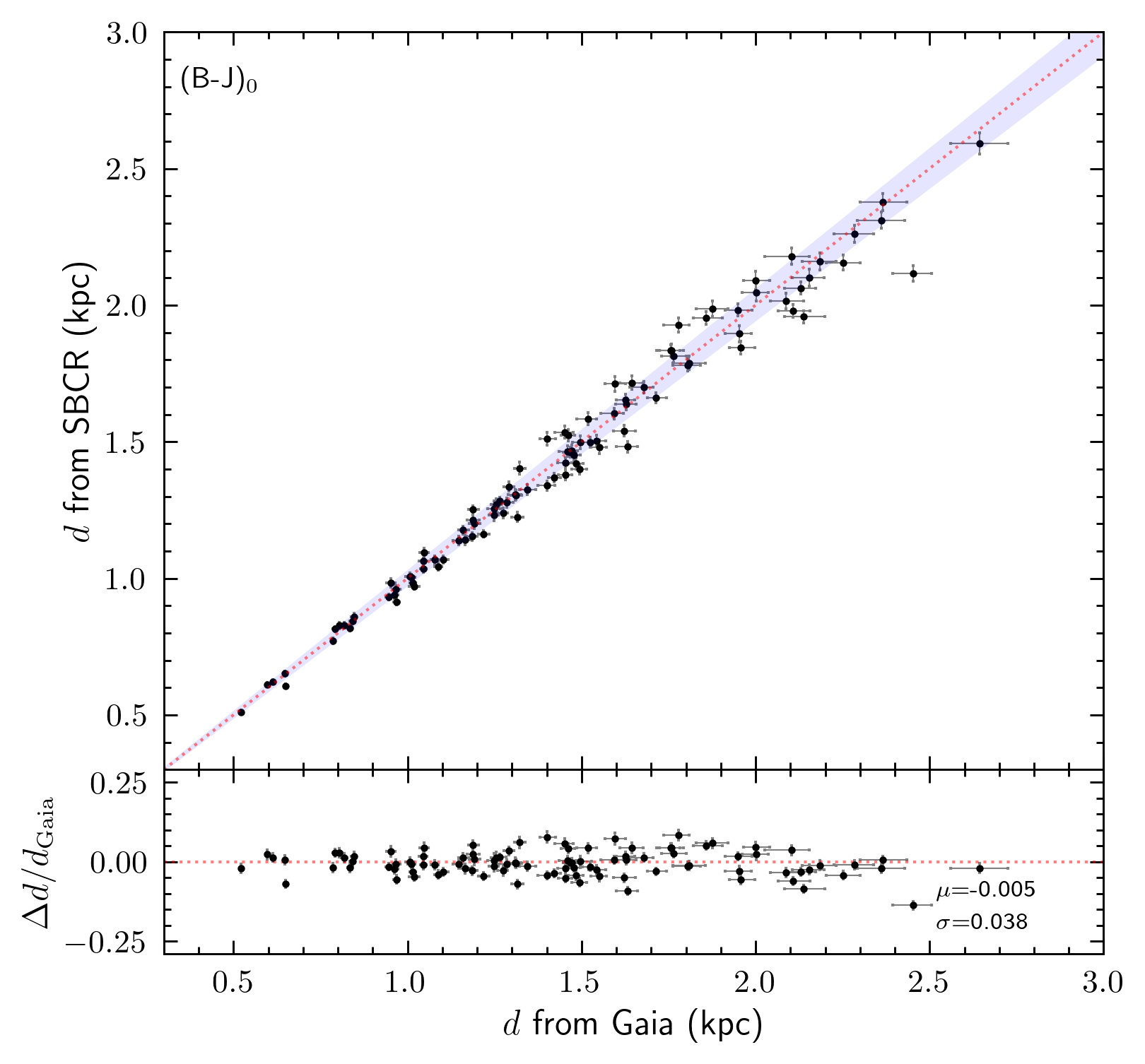}
    \end{subfigure}
    \begin{subfigure}{0.3\textwidth}
        \includegraphics[width=\textwidth]{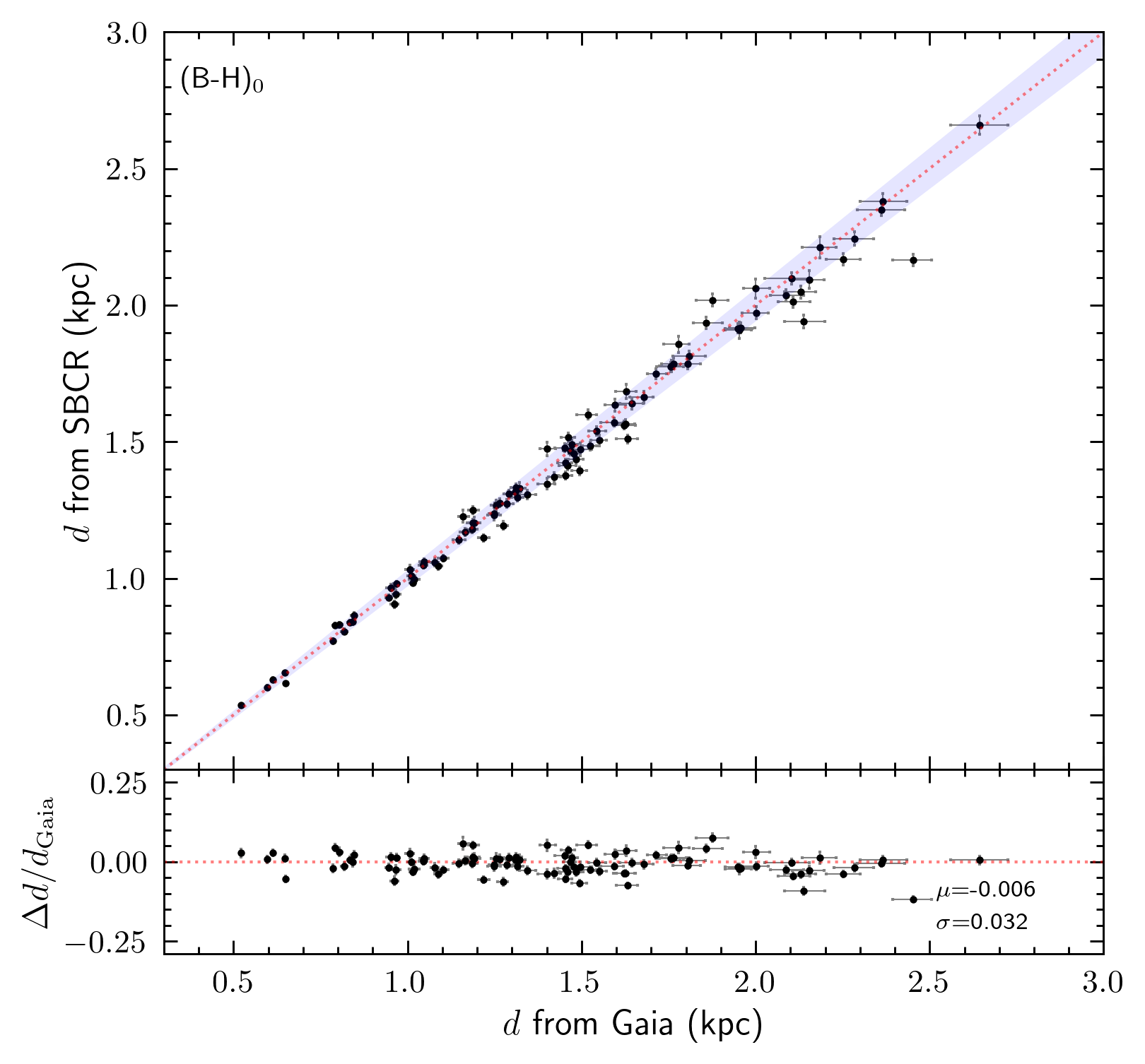}
    \end{subfigure}
    \begin{subfigure}{0.3\textwidth}
        \includegraphics[width=\textwidth]{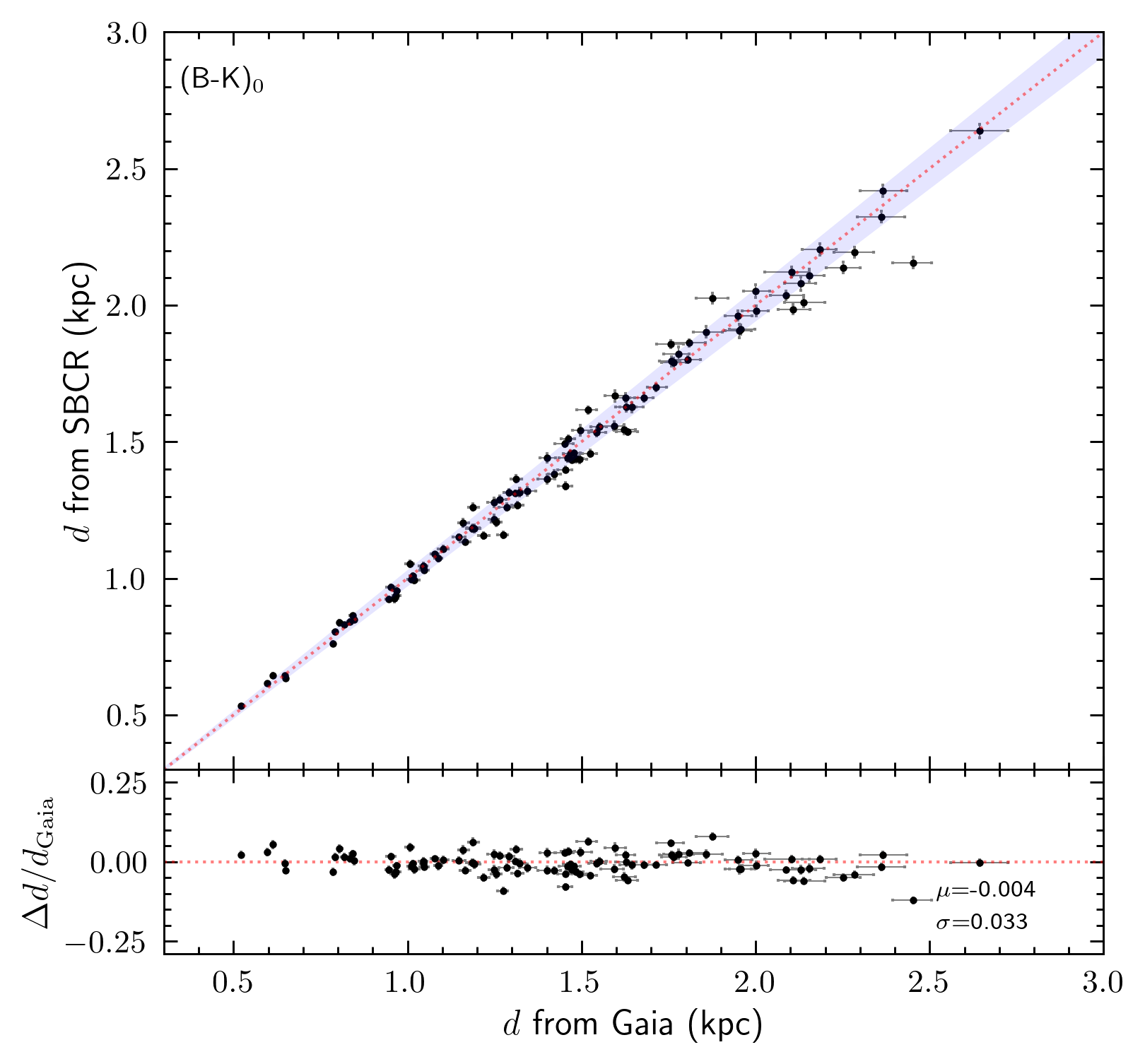}
    \end{subfigure}
    
    \begin{subfigure}{0.3\textwidth}
        \includegraphics[width=\textwidth]{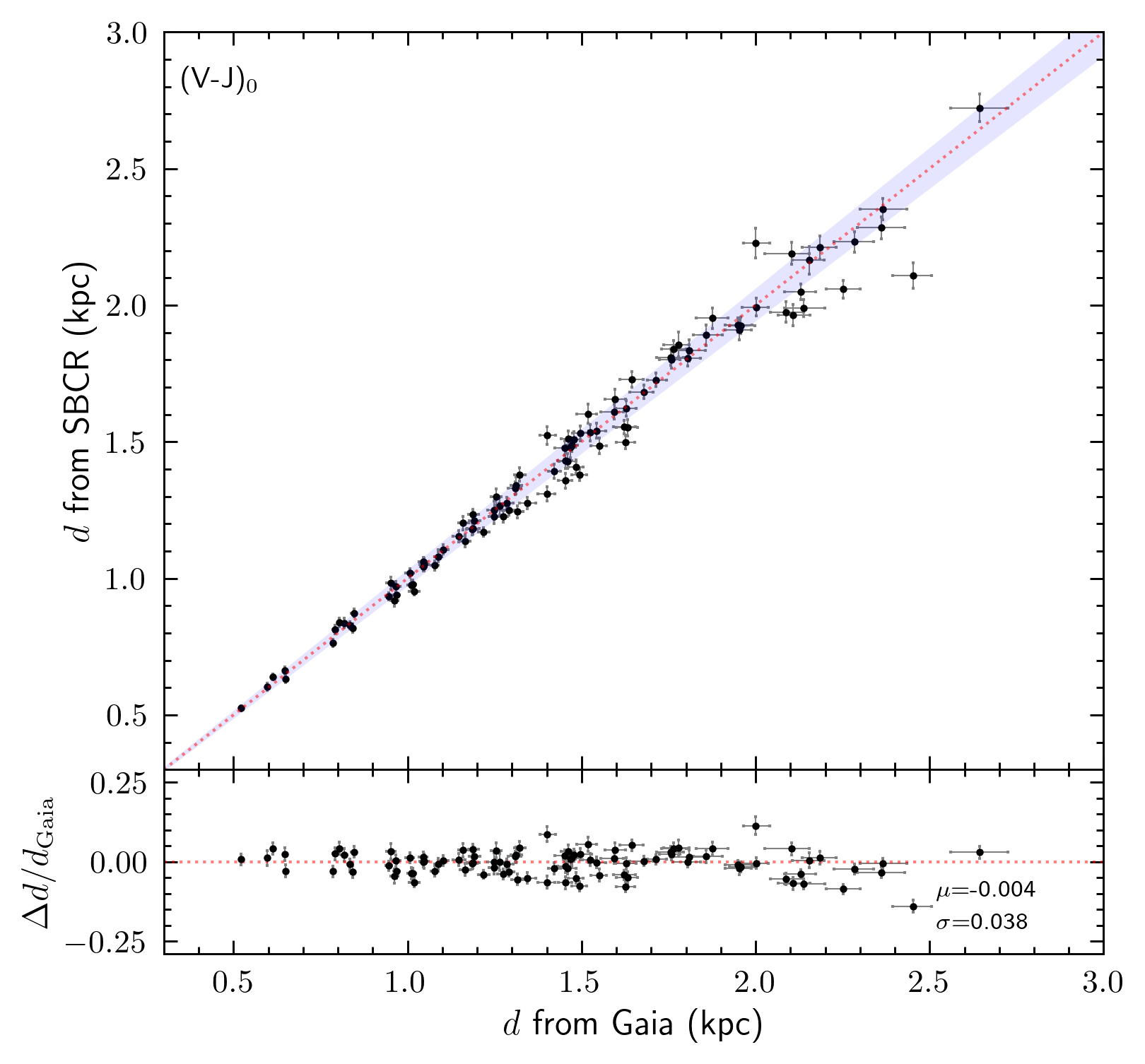}
    \end{subfigure}
    \begin{subfigure}{0.3\textwidth}
        \includegraphics[width=\textwidth]{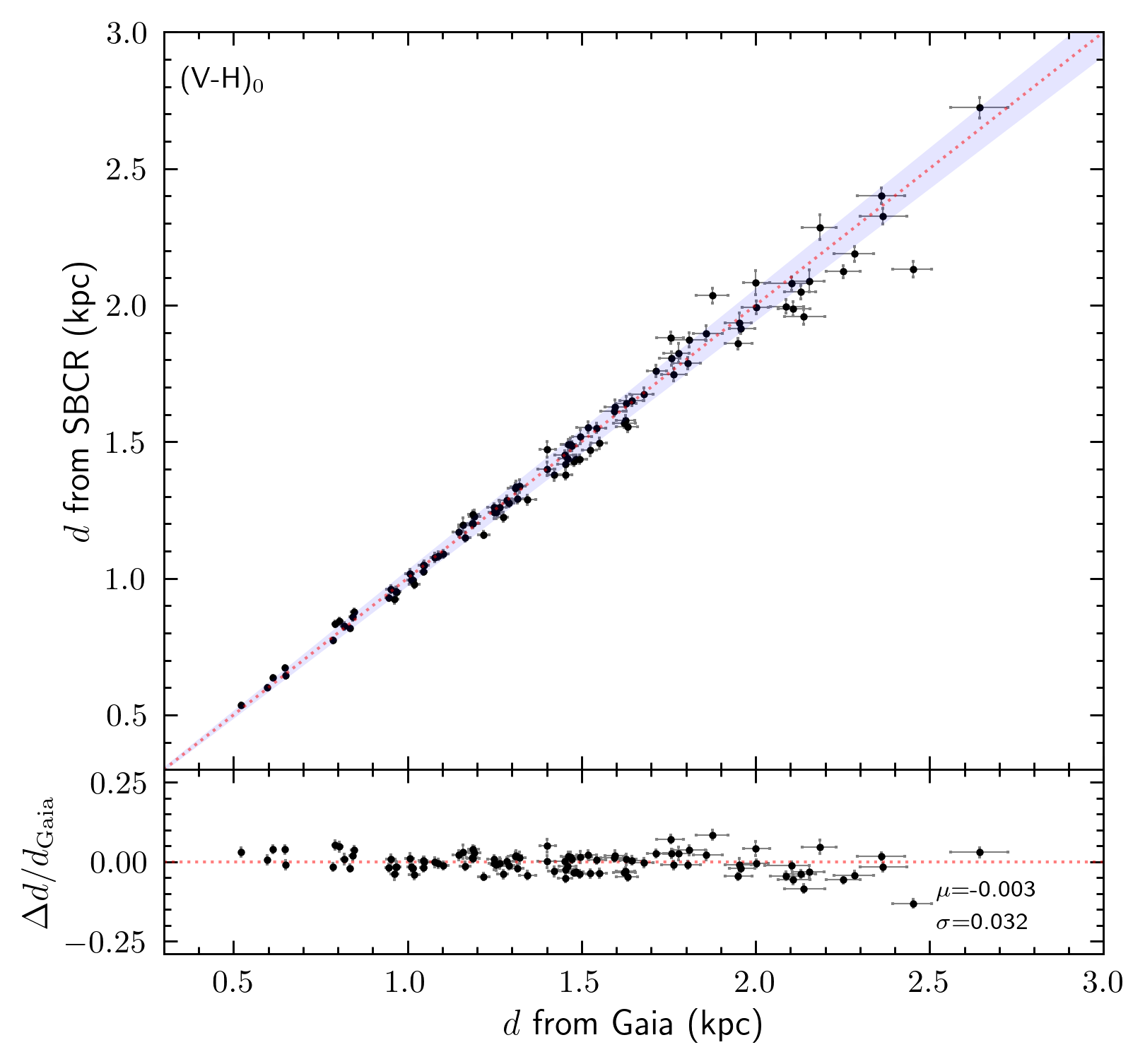}
    \end{subfigure}
    \begin{subfigure}{0.3\textwidth}
        \includegraphics[width=\textwidth]{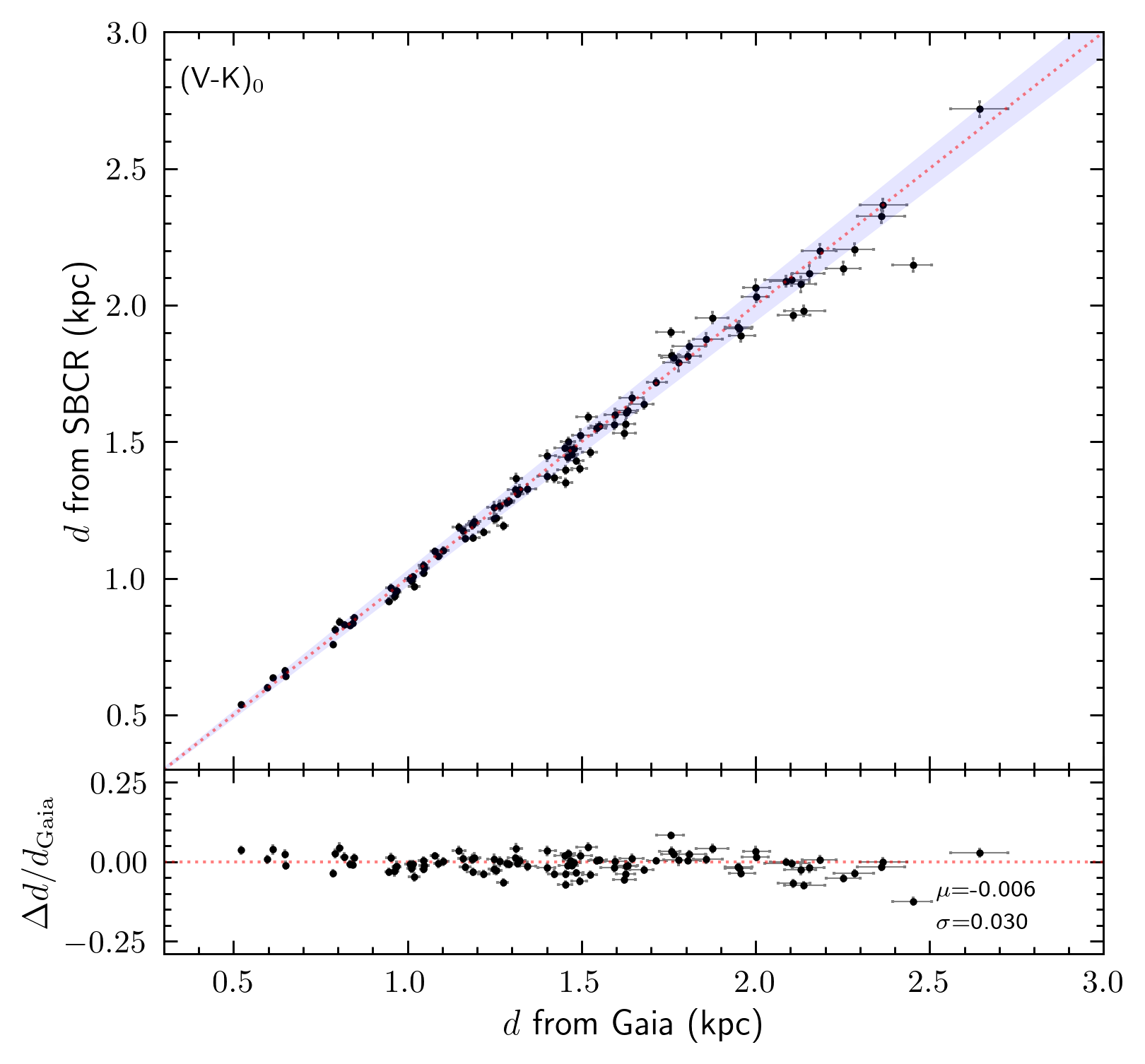}
    \end{subfigure}

    \begin{subfigure}{0.3\textwidth}
        \includegraphics[width=\textwidth]{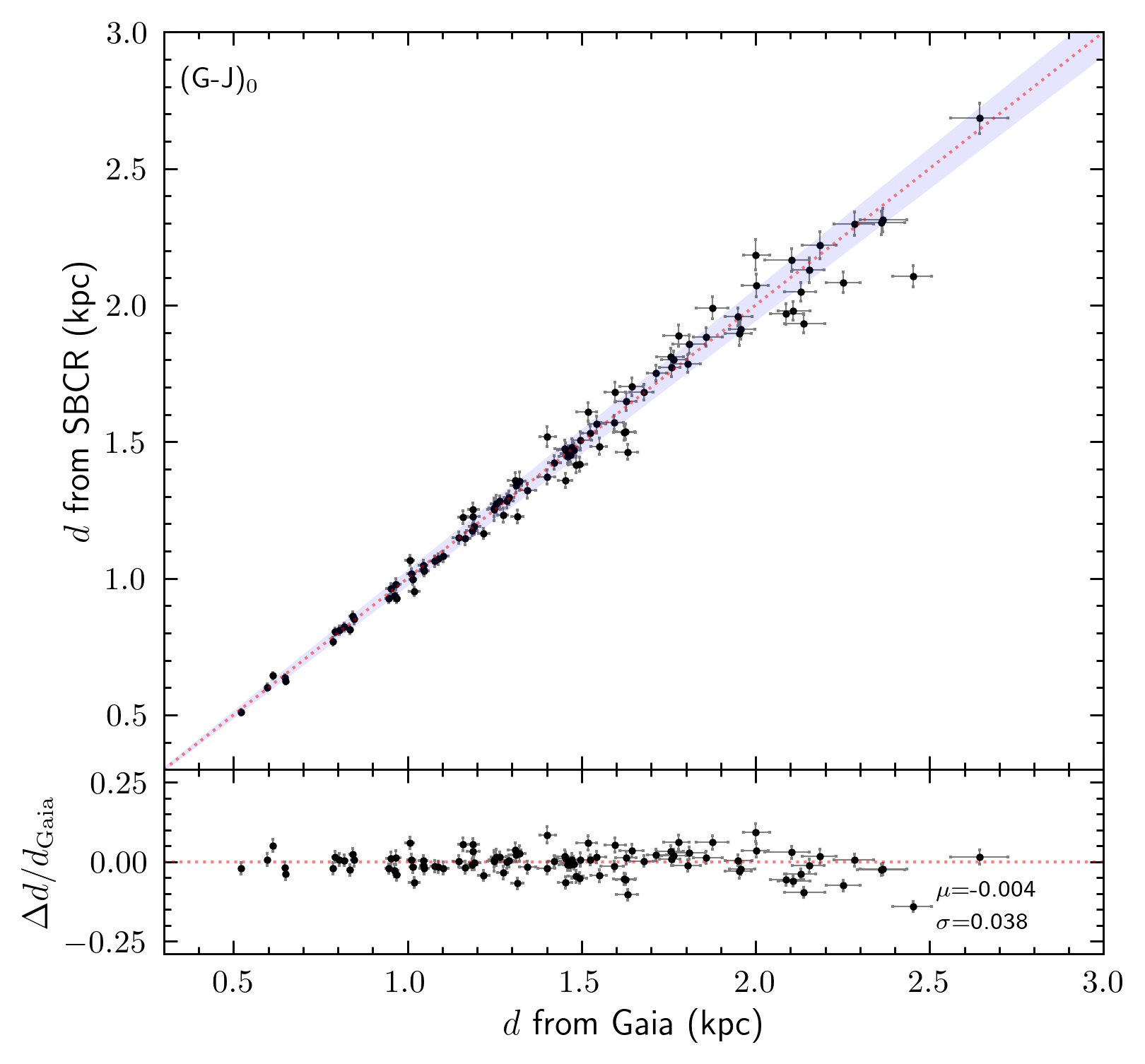}
    \end{subfigure}
    \begin{subfigure}{0.3\textwidth}
        \includegraphics[width=\textwidth]{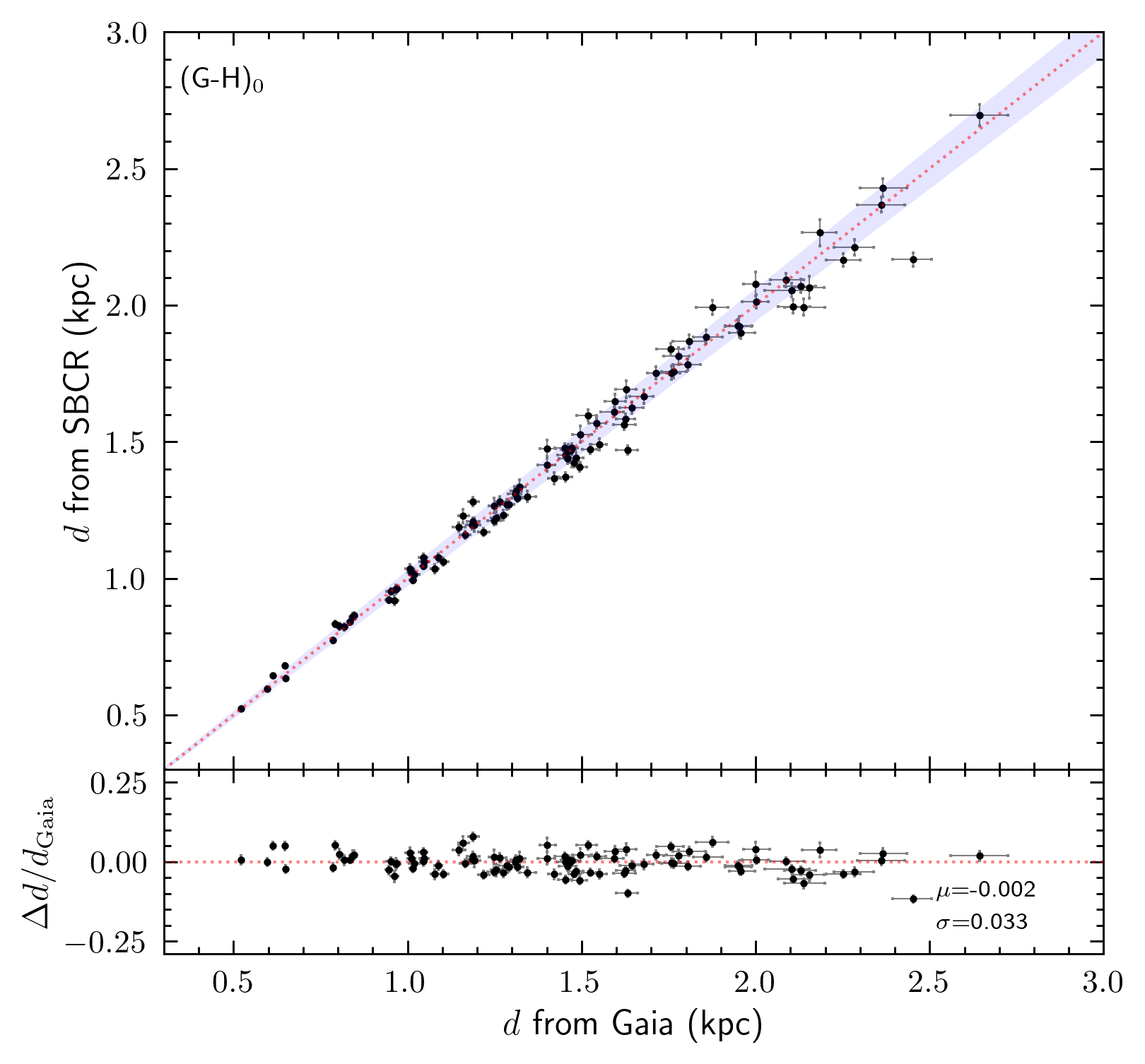}
    \end{subfigure}
    \begin{subfigure}{0.3\textwidth}
        \includegraphics[width=\textwidth]{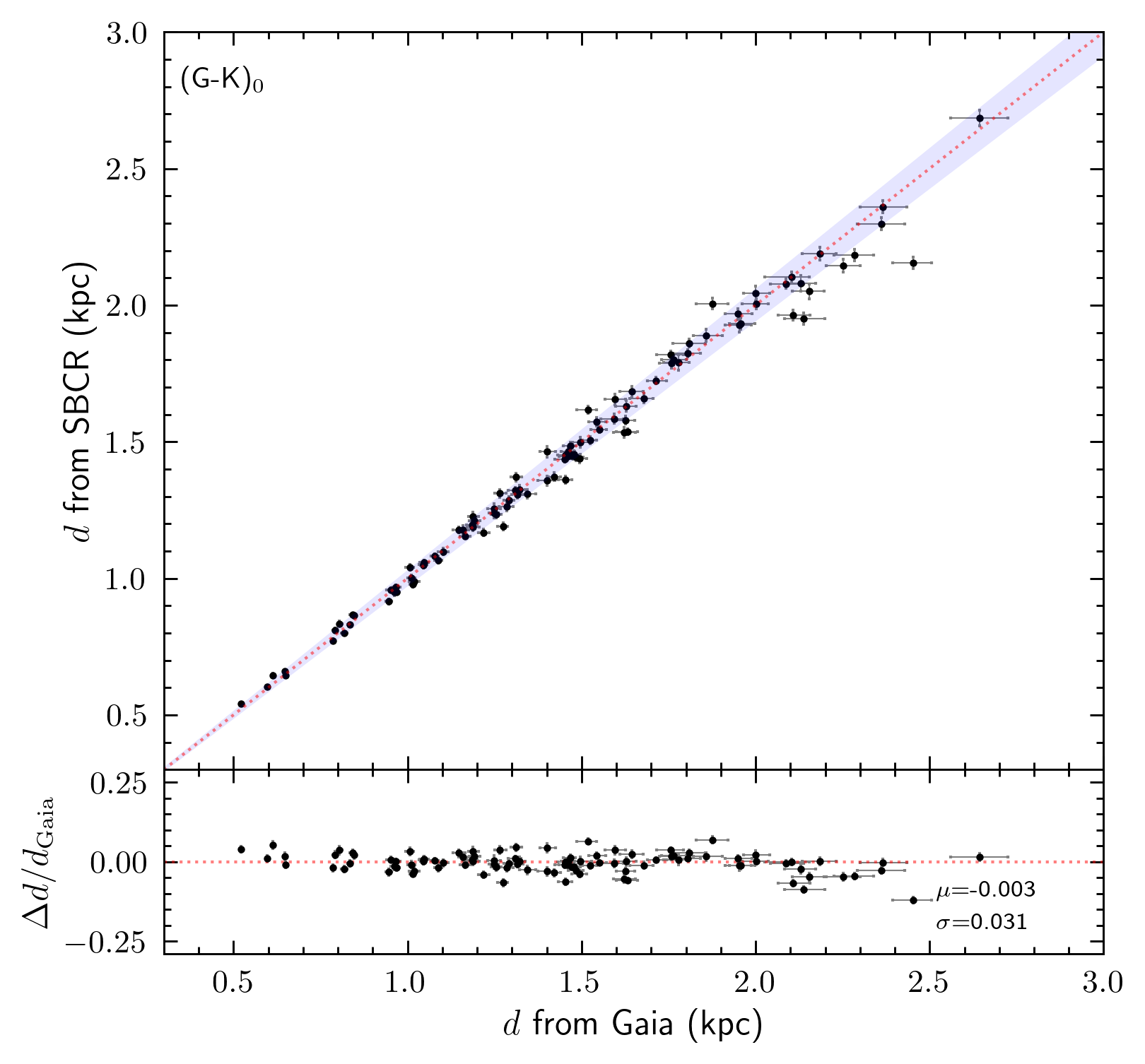}
    \end{subfigure}
    \caption{Validation results of the testing samples. Similar to \ref{fig:validation_fitting}, the $x$-axis represents distances measured by Gaia \citep{2021AJ....161..147B}, while the $y$-axis shows distances estimated using the SBCRs in the Johnson $B$ and $V$ bands and the Gaia $G$ band. The blue shaded area indicates a deviation range of $\pm$3\%. The lower subplot displays the distribution of relative errors ($\Delta d/ d_{Gaia}$, $\Delta d$ is calculated by $d_{SBCR}$-$d_{Gaia}$).}\label{fig:validation_test}
\end{figure*}

Fig.~\ref{fig:validation_test} shows the results of the independent distance calibration performed using the 100 testing samples. These 100 samples are independent of the fitting samples and were not involved in the calibration of the SBCRs. Similar to Fig.~\ref{fig:validation_fitting}, the $x$ axis and $y$ axis represent the geometric distances obtained from Gaia and the distances calculated using our SBCRs, respectively. The bottom panel shows the distribution of relative residuals between the predicted and Gaia distances. 

As shown in Fig.~\ref{fig:validation_test}, for the 100 independent testing samples, it can be seen that for all nine color indices used, the relative differences between the SBCR-derived distances and Gaia's geometric distances are much smaller than their standard deviations, and their weighted mean is $-0.004$. This indicates that there is no systematic bias between the SBCR-derived distances and Gaia's geometric distances. The average dispersion is around 3\%, which corresponds to Gaia's distance uncertainties. In summary, the overall consistency between the SBCR-predicted distances and Gaia's geometric distances validates the accuracy scalability of our relations.

\begin{figure}[h]
    \centering
    \includegraphics[scale=0.55]{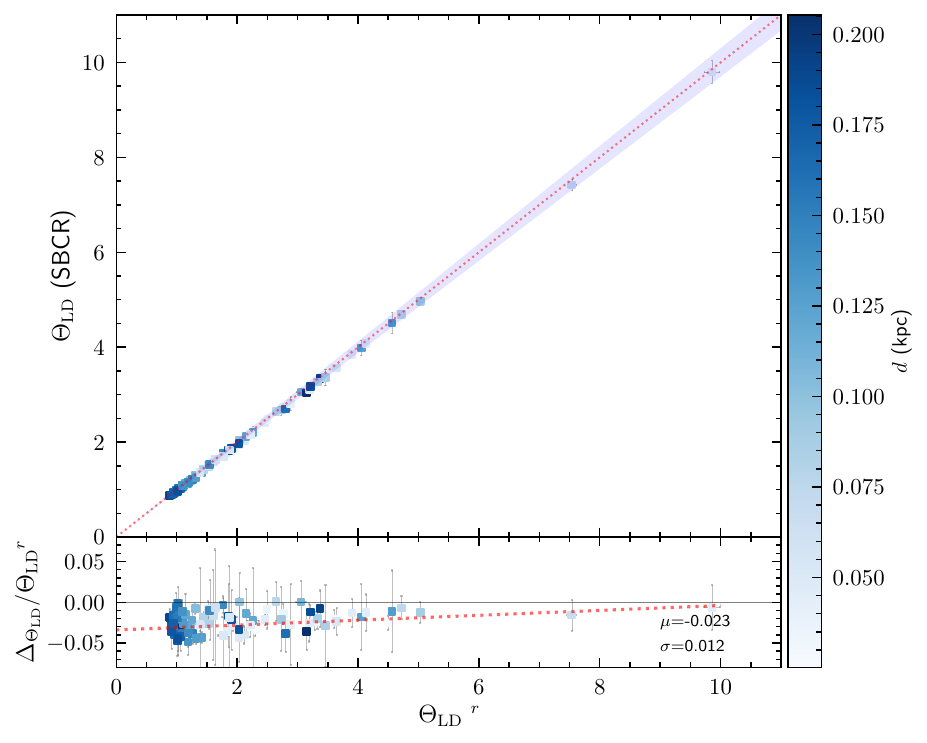}
    \centering
    \caption{Comparison of angular diameters ($\Theta _{\rm{LD}}$). The $x$-axis represents $\Theta _{\rm{LD}}$ directly measured using long-baseline interferometry \citep{2019Natur.567..200P, 2020A&A...640A...2S, 2021A&A...652A..26S}, while the $y$-axis shows angular diameters estimated using the relation proposed in this study (($S_{V}=(1.230\pm0.021)(V-K_{s})_{0}+(2.845\pm0.048)$)). The blue shaded area indicates a deviation range of $\pm$3\%. The bottom panel displays the relative differences between the SBCR-derived angular diameters and the interferometric angular diameters. The colors of the squares represent the distances of target (derived from parallaxes of \textsc{Hipparcos} \citep{2007A&A...474..653V}), darker colors correspond to the larger distances }\label{fig:theta_compare}
\end{figure}

\begin{figure*}[h]
    \centering
    \begin{subfigure}{0.45\textwidth}
        \includegraphics[width=\textwidth]{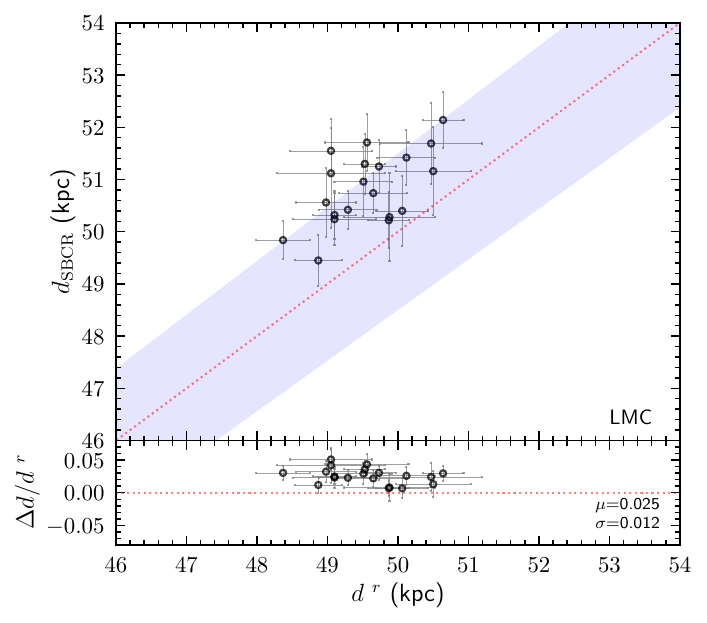}
    \end{subfigure}
    \begin{subfigure}{0.45\textwidth}
        \includegraphics[width=\textwidth]{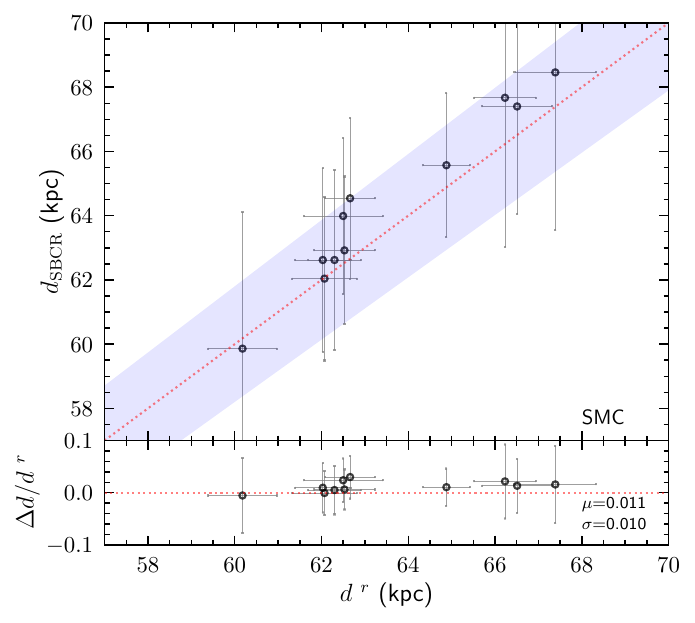}
    \end{subfigure}
    \centering
    \caption{Comparison of distance ($d$) estimates for eclipsing binaries in the LMC (left panel) and the SMC (right panel). The $x$-axis represents distances ($d^r$) of eclipsing binaries derived from the literature ($d^r$). The $y$-axis shows distances ($d_{\rm{SBCR}}$) estimated using the SBCR developed in this study (($S_{V}=(1.230\pm0.021)(V-K_{s})_{0}+(2.845\pm0.048)$)). The blue shaded area indicates a deviation range of $\pm$3\%. The bottom panel illustrates the distribution of relative errors ($\Delta d/d^r = (d_{\rm{SBCR}}-d^r)/d^r$).}\label{fig:LMC_compare}
\end{figure*}

\subsection{External validation}
\subsubsection{Comparison of angular diameters}
For the external validation, we compile a sample of 109 RGB and RC stars from the literature \citep{2019Natur.567..200P, 2020A&A...640A...2S} (the open squares as shown in Fig.~\ref{fig:VK_SBCR_compare}), all of which have angular diameters measured through infrared interferometry. These angular diameters were obtained using the ESO VLTI and PIONIER instruments, or compiled from the JMDC, achieving a precision of 1\%. The JMDC is the most comprehensive and up-to-date catalog listing all interferometric measurements. We used angular diameters ($\theta_{\rm{LD}}$) compiled from the literature as reference measurements. Using the SBCR in $V-K$ color, from Eq.~\ref{Surface}, the $\theta_{\rm{LD}}$ can be derived by Eq.~\ref{theta_Sv}:

\begin{equation}
   \theta_{\rm{LD}} = 10^{0.2(S_{V}-m_{V_{0})}} \label{theta_Sv}
\end{equation} 

Here, $S_{V}$ is derived from the SBCR of $V-K$ color. The magnituides in the $V$, $K$ bands and the corresponding extinctions are also sourced from \citet{2019Natur.567..200P} and \citet{2020A&A...640A...2S}.

Fig.~\ref{fig:theta_compare} shows a comparison of angular diameters, where the horizontal axis represents the angular diameters obtained through infrared interferometry as reported in the literature, and the vertical axis represents the angular diameters estimated using our SBCR, and the blue-shaded area indicates a deviation range of $\pm$3\%. The bottom panel illustrates the relative differences between the SBCR-derived angular diameters and the directly observed angular diameters. It is evident that, compared to direct infrared interferometry measurements, angular diameters derived by our SBCR exhibit a systematic bias of approximately -2.3\%. The uncertainties and slight systematic bias arise from a complex interplay of factors, including asteroseismology, astrometry, and interferometry. 

As demonstrated in Eq.~\ref{theta}, the angular diameter is associated with both the radius and the distance. Consequently, we first obtain the distances of these targets, which are shown by color in Fig.\ref{fig:theta_compare}, with darker colors representing greater distances. The distances depicted in Fig.\ref{fig:theta_compare} have been derived from the parallaxes of \textsc{Hipparcos} \citep{2007A&A...474..653V}. Although Gaia provides more precise parallaxes, but for sources with G $<$ 6 \texttt{mag} (these targets are very bright (G $\sim$5 \texttt{mag})), it is evident that the uncalibrated CCD saturation might lead to unreliable astrometric solutions \citep{2018A&A...616A...2L}, and in this case, \textsc{Hipparcos} might be more reliable \citep{2020A&A...643A.116D}. A comparison of the parallaxes of these sources from Gaia (RUWE $<$ 1.4, 40 stars) and \textsc{Hipparcos} was also conducted, the results indicate that \textsc{Hipparcos} parallaxes tend to overestimate the parallaxes by approximately 0.244 mas in comparison to Gaia. As shown in Fig.\ref{fig:theta_compare}, all of these interferometric targets are within 200 \texttt{pc}. Furthermore, it also indicates that the relative difference in angular diameters is getting smaller when the angular diameter is getting larger. Here, due to the lack of interferometric data for more targets, we are unable to quantitatively confirm the discrepancies caused by distance.

In addition to the distance,
the radius should also be discussed. In this paper, we adopt the radius provided by \citet{2023ApJ...953..182W}. They used a grid-based asteroseismic modeling approach (the grid was calculated by \citet{2022ApJ...927..167L}) and incorporated two additional observed constraints (e.g. the gravity-mode period spacing and luminosity) to measure the radii, which significantly improved the accuracy of the radii. However, the inclusion of these constraints may introduce systematic bias (see Table 2 in \citet{2023ApJ...953..182W}). Additionally, the accuracy here only refers to the measurement uncertainties of the parameters compared to the model. Whether there are systematic biases requires more external data to analyze. 

Moreover, by combining with Gaia DR3 parallaxes, \citet{2024A&A...690A.327V} compared asteroseismic radii with those estimated using two SBCRs presented by \citet{2019Natur.567..200P} and \citet{2021A&A...652A..26S}. These two SBCRs are established by the samples with interferometric angular diameters. They found that, compared to the asteroseismic radii, the stellar radii estimated using the SBCR from \citet{2019Natur.567..200P} were overestimated by 1.2\%, whereas those estimated using the SBCR from \citet{2021A&A...652A..26S} were underestimated by 2.5\%. They also indicated that the deviations are related to parallax and [$\alpha/$Fe], and the different deviations are observed on the RC stars and RGB stars, as well as different deviations in the K2 and Kepler data. In addition, discussions in the measurement of radii obtained from asteroseismology and interferometric measurements have been reported by various authors (see, e.g. \citet{2020FrASS...7....3H, 2022MNRAS.517.4187T}). This suggests that there are small but non-negligible differences between the radii measured by asteroseismology and those measured by infrared interferometry. However, the sample size and homogeneity of the data limit our understanding of the systematics in asteroseismic radii.

\subsubsection{Comparison of distances}

Based on eclipsing binary systems with precise radius measurements in the LMC and SMC, SBCRs were used to measure the distances for the LMC and SMC \citep{2019Natur.567..200P, 2020ApJ...904...13G}. Therefore, similar to \citet{2019Natur.567..200P} and \citet{2020ApJ...904...13G}, we apply our SBCR (($S_{V}=(1.230\pm0.021)(V-K_{s})_{0}+(2.845\pm0.048)$)) to measure the distances of the eclipsing binaries in LMC and SMC. In these two works, the distances were estimated by the SBCR in \citet{2019Natur.567..200P} ($S_{V}=(1.330\pm0.017)[(V-K)_{0}-2.405]+(5.869\pm0.003)$).

Fig.\ref{fig:LMC_compare} presents a comparison between the distances of the detached eclipsing binary samples in the LMC (left panel) and the SMC (right panel). In Fig.~\ref{fig:LMC_compare}, the $x$-axis represents the distances derived from the literature ($d^r$). The $y$-axis shows the distances derived using our relation ($d_{\rm{SBCR}}$). The bottom panel illustrates the distribution of relative residuals ($\Delta d/d^r = (d_{\rm{SBCR}}-d^r)/d^r$) between the distances we measured and those obtained from the literature. These binaries are the same sources as
those used by \citet{2019Natur.567..200P} and \citet{2020ApJ...904...13G}. From their samples, we extracted the radii, $(V-K)_{0}$ values, and $V_{0}$ magnitudes for the eclipsing binaries. For each eclipsing binary, we obtain the mean value of the distances for two components as the final distance of each binary.

As shown in Fig.\ref{fig:LMC_compare}, for the binaries in the LMC and SMC, the distances predicted using our SBCR are in good agreement with those reported in the literature (dispersion $\sim$ 1\%). And the slight overestimations of 2.5\% and 1.1\% are shown. These overestimations may arise from the systematic underestimation of the angular diameters obtained using our SBCR compared to infrared interferometric measurements. In this paper, we do not apply any corrections for this deviation here.

\section{Discussion and conclusions} \label{summary}

In this paper, we have established the surface brightness-color relation by using 626 red giant branch stars with high-precision asteroseismic radii and Gaia photometry, and validated this relation using Gaia-independent distances. Additionally, we have further tested the relation through the eclipsing binary distance measurements in the Large Magellanic Cloud and Small Magellanic Cloud. Based on the 526 RGB stars with high-precision asteroseismic radii (precision $\sim$ 1\%) and Gaia's distances (uncertainties $<$ 3.5\%), we developed nine SBCRs for giants in the Johnson $B$, Johnson $V$, and Gaia $G$ bands, in combination with 2MASS $J$, $H$, and $K$ bands. \citet{2021A&A...649A.109G} previously established SBCRs for dwarfs and subgiants within these same bands. Theoretical analysis have demonstrated that SBCRs are influenced by different spectral types and luminosity classes \citep{2022A&A...662A.120S}. Our work uniquely focuses on giant stars, thereby filling a critical gap in the existing literature. This expansion not only broadens the applicability of SBCRs but also facilitates more accurate distance measurements across a wider variety of stellar types.

Analysis of the test data indicated that the distances derived from the SBCRs across various color indices are consistent with Gaia's distances, exhibiting a dispersion that increases with distance and averages around 3\% to 4\%. Compared to the angular diameters measured by infrared interferometry, the angular diameters derived from our SBCRs were underestimated by approximately 2.3\%. Moreover, the relative difference in angular diameter decreases with increasing angular diameter. These biases may arise from asteroseismic radii, astrometric measurements and interferometric data. Additionally, distances of eclipsing binaries in the Large Magellanic Cloud and Small Magellanic Cloud obtained using our SBCRs are generally consistent with those measured through other relations, showing a dispersion of approximately 1\%, but with a slight overestimation of 1\% $\sim$ 2.5\%. The consistency between the two methods requires further verification with additional data.

The key strength of our study lies in the large and homogeneous samples used to construct the SBCRs, as well as the use of more distant samples observed by Gaia for independent distance validation. In previous studies, the number of samples used to establish and validate the SBCR based on direct measurements was small (approximately 109 RGB and RC stars, as summarized by \citet{2019Natur.567..200P} and \citet{2020A&A...640A...2S} ) and these samples are the nearby targets (within $\sim$200 \texttt{pc}). In this study, we used Gaia-independent distance to demonstrate the effectiveness of SBCR for more distant targets. Compared to the observations of infrared interferometry, time-domain spectroscopic and photometric data, we utilized a substantially larger sample, required lower observational costs to obtain sample parameters, and employed a unified methodology for measuring angular diameters.

For further calibration, the advancement of comprehensive photometric surveys allows the effective extension of target samples with high quality radius and distance measurements. The integration of data from missions such as TESS and Gaia will provide comprehensive coverage and consistency of photometric measurements, thereby reducing uncertainties and improving the robustness of surface brightness-color relations across different stellar populations.

\begin{acknowledgements}
We wish to thank the referee for his/her valuable comments and suggestions, which have helped us further improve this work. This work is supported by the National Natural Science Foundation of China (NSFC) with grant Nos.12288102, 12125303, 12090040/3, the National Key R\&D Program of China (grant No.2021YFA1600401/ 2021YFA1600403), the NFSC (grant Nos.12303106, 12473034, 12103086, 12373037, 12173047), the Postdoctoral Fellowship Program of CPSF (No.GZC20232976), Yunnan Fundamental Research Projects (grant Nos.202401AT070139, 202101AU070276), the International Centre of Supernovae, Yunnan Key Laboratory (No.202302AN360001) and the Yunnan Revitalization Talent Support Program—Science \& Technology Champion Project (No.202305AB350003). This work is also supported by the China Manned Space Project of No.CMS-CSST-2021-A10. We thank the Gaia Data Processing and Analysis Consortium (DPAC) for their substantial contributions in producing and releasing high-quality data.
\end{acknowledgements}

%-------------------------------------------------------------------
\bibliographystyle{aa}
\bibliography{star}

\end{CJK}
\end{document}